\documentclass[sigconf,10pt]{acmart}
\usepackage{algorithm}
\usepackage{algorithmicx}
\usepackage[noend]{algpseudocode}
\usepackage{amsmath}
\usepackage{amssymb}
\usepackage{accents}
\usepackage{xcolor}
\usepackage{xspace}
\usepackage{balance}
\usepackage{enumitem}
\usepackage[ampersand]{easylist}
\usepackage{pifont}
\usepackage[utf8]{inputenc} 
\usepackage{pbox}
\usepackage{epstopdf}
\usepackage[ampersand]{easylist}
\usepackage{multirow}
\usepackage{graphicx}
\usepackage[caption=false]{subfig}
\usepackage{etoolbox}
\patchcmd{\quote}{\rightmargin}{\leftmargin 1em \rightmargin}{}{}
\usepackage{local} 
\usepackage{soul}

\usepackage{url} 
\makeatletter
\g@addto@macro{\UrlBreaks}{\UrlOrds}
\makeatother

\renewcommand\footnotetextcopyrightpermission[1]{} 
\newif\iffull
\fullfalse 

\renewcommand{\hl}[1]{#1} 
\renewcommand{\hlmath}[1]{#1}

\begin{document}
\title{Control Plane Compression}
\subtitle{Extended Version of the SIGCOMM 2018 Paper}

\author{Ryan Beckett}
\affiliation{
  \institution{Princeton University}
}
%
\author{Aarti Gupta}
\affiliation{
  \institution{Princeton University}
}
%
\author{Ratul Mahajan}
\affiliation{
  \institution{Intentionet}
}
%
\author{David Walker}
 \affiliation{
  \institution{Princeton University}
 }

\maketitle

%
%
%
%

\textbf{Abstract---}
We develop an algorithm capable of compressing large networks into a smaller ones with similar control plane behavior: For every stable routing solution in the large, original network, there exists a corresponding solution in the compressed network, and vice versa. Our compression algorithm preserves a wide variety of network properties including reachability, loop freedom, and path length. Consequently, operators may speed up network analysis, based on simulation, emulation, or verification, by analyzing only the compressed network. Our approach is based on a new theory of control plane equivalence. We implement these ideas in a tool called \sysname and apply it to real and synthetic networks.
Bonsai can shrink real networks by over a factor of 5 and speed up analysis by several orders of magnitude.

%
%
%
%

\section{Introduction} \label{sec:introduction}

Configuration errors are a leading cause of network outages and security breaches~\cite{juniper-study, yankee-study, summer-of-misconfiguration, azure-outage, bgp-misconfiguration, time-warner}. For instance, a recent misconfiguration disrupted Internet connectivity for millions of users in the USA for over 1.5 hours, and similar incidents last year impacted users in Japan, India, Brazil, Azerbaijan, and beyond~\cite{bgpmon}. 

The root cause of many of these errors is simply complexity:
Networks typically run one or more distributed routing protocols that exchange information about available paths to destinations. The paths that are advertised and preferred by routers are determined by their configuration files, which can easily contain tens of thousands of low-level, vendor-specific primitives. Unfortunately, it is nearly impossible for human network operators to reason about the correctness of these files, let alone that of behavior that results from distributed interactions between many routers.

To minimize errors in configurations, researchers have taken
a number of approaches to find bugs and check correctness, including static analysis~\cite{rcc, minerals}, simulation~\cite{batfish, bgp-prediction}, emulation~\cite{crystalnet}, monitoring~\cite{veriflow, netplumber, delta-net, atomic}, model checking~\cite{plankton}, systematic testing~\cite{era, symnet}, and verification~\cite{arc,bagpipe,minesweeper}.
%
Yet for almost all such techniques, scaling to large networks remains challenging. 
For example, in Batfish~\cite{batfish},  a testing tool, 
the time it takes to model control plane dynamics
limits the number of tests that can be administered.
Similarly, the cost of the verification tool Minesweeper~\cite{minesweeper}
grows exponentially in the worst case, and in practice, tops out at
a few hundred devices---far short of the 1000+ devices that
are used to operate many modern data centers.




In this paper, we tackle these problems by defining a new theory of
\emph{control plane equivalence} and using it to compress large,
concrete networks into smaller, abstract networks with equivalent 
behavior. Because our compression techniques
preserve many properties of the network control
plane, including reachability,
path length, and loop freedom, analysis tools of all kinds can operate (quickly)
on the smaller
networks, rather than their large concrete counterparts.
In other words, this theory is an effective complement to 
ongoing work on network analysis, capable of helping accelerate
a wide variety of analysis tools.  Moreover,
because our transformations are bisimulations, rather than
over- or under-approximations, tools built on our theory can avoid
both unsound inferences and false positives.

Intuitively, the reason it is possible to compress control planes in this
fashion is that large networks tend to contain quite a bit of structural symmetry---if not, they
would be even harder to manage. For instance, many spine (or leaf or aggregation) routers in a data center may be similarly configured; and, as we show later, symmetries exist in backbone network as well.
%
Recently, Plotkin \emph{et al.}~\cite{surgery}  exploited similar intuition to develop other tools for network verification. However, they operate over the (stateless) network data plane, \emph{i.e.,} the packet-forwarding rules, whereas we operate over the control plane, \emph{i.e.,} the protocols that distribute the available routes.  While both the data and control planes process messages (data packets and routing messages, respectively), the routing messages interact with one another whereas the data packets do not. More specifically, data packet processing depends only on the static packet-forwarding rules of a router; it does not depend on other data packets. In contrast, routing messages interact:  the presence and timing of one (more preferred) message can cause another (less preferred) message to be ignored. 
%
%
\hl{Such interactions create dynamics not present in stateless
data planes and can even lead to many different routing solutions for the same network. In other work,
Wang} \ETAL~\cite{bgp-compression} \hl{also observed that compression can lead to improved control plane analysis
  performance, and they designed an algorithm for compressing BGP networks
  that can speed analysis of convergence
  behavior.  In contrast, our algorithms are designed to preserve arbitrary path properties of
  networks.  Such properties include reachability, loop freedom, absence of black holes and access
  control.  We also define our algorithms 
  over a generic protocol model so we can apply the ideas to a wide range of protocols ranging from
  BGP to OSPF to static routes.}

More specifically, we make two core contributions:

\para{A Theory of Control Plane Equivalence}
Our theoretical development begins by defining the
Stable Routing Problem (SRP), a generic model of a routing
protocol and the network on which it runs.  SRPs can model
networks running a wide variety of  protocols including
distance-vector, link-state, and path-vector
protocols.  SRPs are directly inspired by the stable paths
problems (SPP)~\cite{stable-paths}, but rather than describing the
protocols' final solution using end-to-end paths (as SPPs do), SRPs
describe runtime routing behavior in terms of local processing
of routing messages, as configurations do.
In addition to modeling raw configurations more closely,
this distinction allows SRPs to capture a wider
variety of routing behaviors that emerge at runtime, including static routing and
other protocols that may generate loops. Consequently, our formulation of SRPs is similar to the protocol models used by routing algebras~\cite{routingalgebra,metarouting}, though this earlier work focused on convergence rather than network compression or analysis of topologically-sensitive properties such as reachability.


With a network model in hand, we turn to the process of
characterizing network transformations.
Intuitively, we would like to define transformations that convert
concrete networks into abstract ones that make equivalent
control decisions and generate similar forwarding behavior.  However,
doing so directly is challenging as validating that two
SRPs are equivalent is 
as hard as 
the control plane verification problem
we are trying to speed up in the first place!
We address this challenge by defining a class of network transformations
that we call \emph{effective abstractions}.  These abstractions
are characterized by conditions designed to be checked efficiently,
and locally, without the need for a global simulation.
Our central theoretical result is that these conditions imply behavioral
equivalence of the concrete and abstract networks.

\para{An Efficient Compression Algorithm}
Our theory paves the way for a practical algorithm for automatically computing
compact, abstract control planes from configurations of large networks.
The algorithm is based on abstraction refinement: Starting with the
coarsest possible abstraction
it iteratively refines the abstract network, checking at
each step to determine whether or not it has found an
effective abstraction. To implement such checks efficiently,
we use a variety of efficient data structures such as Binary Decision 
Diagrams to represent configuration semantics.
%
\hl{In practice, 
  the algorithm reduces network sizes significantly, bringing more networks into range
  for various analyses.  For example, our tool was able to compress an operational 196-node data center network down to 26 nodes and to reduce the number of edges by a factor of roughly 100.  A 1086-node WAN using eBGP, iBGP, OSPF and static
routes was compressed down to 137 nodes and its edge count was reduced by a factor 7.}

\OMIT{
\section{Motivation}
\label{sec:motivation}

Our work is motivated by the observation that most large networks have inherent structural symmetry, and we can speed control plane analysis by systematically exploiting these symmetries. In particular, we want to derive a smaller version of the input network which retains the essential properties of the input network, such that analysis results on the smaller network can be translated directly to those on the original.

For instance, in a large data center network with a standard three-tier Clos topology, many top-of-rack routers (and, respectively, aggregation routers and spine routers) tend to be similarly configured and forward packets similarly. In the extreme, we may be able to shrink such networks down to three routers (i.e., one top-of-rack, one aggregation, and one spine router) for the purposes of the analysis. Of course, whether that is possible depends on how similar the configurations are and what analysis we want to perform (e.g., the number of faults that the smaller 3-node network can tolerate will be different from the original). Our work formally captures what it means for the smaller network to be equivalent, what properties yield identical results for both networks, and presents an algorithm to derive the smaller network.

Plotkin~\ETAL made a similar observation about symmetries in large networks and exploited them for data plane analysis. However, by virtue of operating over the forwarding state of the routers, miniaturization for data plane analysis is qualitatively different from doing so for control plane (i.e., configurations) analysis. The primary difference is that the data plane state directly describes the network's runtime behavior, whose equivalence we want to preserve during miniaturization.  On the other hand, reasoning about the runtime behavior from control plane configuration is a significant challenge by itself~\cite{batfish,minesweeper}----based on their configuration, devices interact in complex ways such that routers with identical configuration can have different runtime behaviors and even multiple different runtime behaviors may emerge for a given configuration. The challenge is compounded when we want to reason not only about runtime behavior of a large network but also transform the  network to a smaller network with similar runtime behaviors. 

\begin{figure}
  \includegraphics[width=\columnwidth]{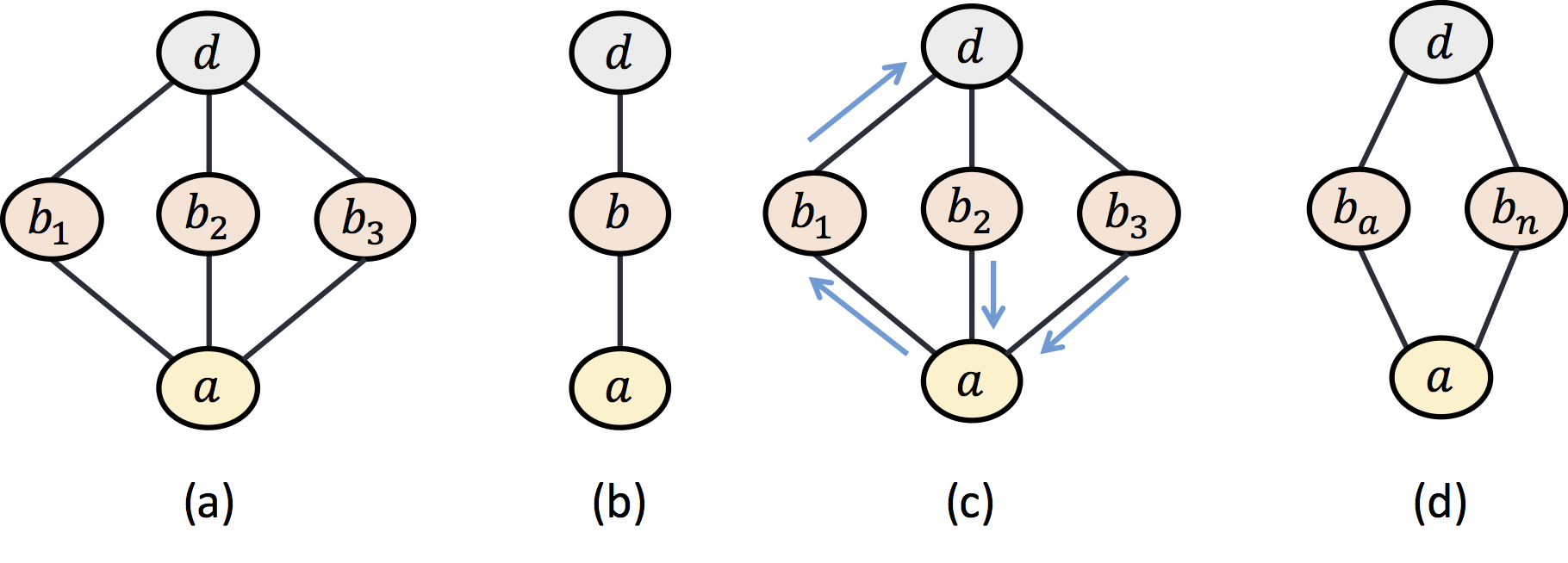}
  \caption{(a) An example network running BGP, where all nodes prefer shortest paths except that $b_i$ routers prefer to go through $a$ when possible. (b) A smaller network that preserves many properties if $b_i$ routers had similar data plane state but does not do that when they have similar configuration. (c) One of three possible forwarding behaviors than can emerge for the network. (d) A smaller network that preserves many properties for the original network.} 
  \label{fig:example}
\end{figure} 

We illustrate the challenges that underlie miniaturization for control plane analysis using the example in Figure~\ref{fig:example}(a). Suppose all routers run eBGP with their neighbors and have default policy (i.e., prefer shorter paths), except that the middle tier of three $b_i$ routers prefer to go through $a$. Because the middle tier of routers have similar policy, one might think that we can miniaturize the network to that in Figure~\ref{fig:example}(b), where the three $b_i$ routers have been combined into a single abstract router. While that may be appropriate if the $b_i$ routers had similar data plane state, this miniaturization is incorrect for control plane analysis. 

The catch is that for destinations advertised by $d$, one of three different runtime behaviors can emerge. One such behavior is shown in Figure~\ref{fig:example}(c), where $b1$ becomes the "anchor" that does not get its preferred path of going through $a$ but enables the other two routers to get their preferred paths. The other two behaviors correspond to $b2$ and $b3$ becoming anchor nodes. Which  of these behaviors will emerge in practice depends on the ordering of routing messages and the tie-breaking policy at $a$ (which dictates route selection among multiple equally desirable options). For simplicity, lets assume that $a$ selects the first route that it gets. Then, the $b_i$ node that first advertises the path to $a$ becomes the anchor because $a$ will then propagate its choice to other $b_i$ nodes, who will prefer going through $a$ instead of using the direct path to $d$. 

At this point, one may conclude that no miniaturization is possible in this network since there is at least one runtime behavior in which a $b_i$ node plays a unique role, which indicates that none of the nodes can be combined. But we describe later how our techniques enable miniaturization even in this setting, by reducing the three $b_i$ routers into two abstract routers, one for the anchor node and one for the non-anchor nodes.  We also describe how our abstraction preserves many properties of interest that go well beyond reachability. The data plane based abstraction of Plotkin \ETAL was limited to reachability analysis. 

\ratul{Talk about structure in a backbone network}
}
%
%
%
%

\section{Overview} \label{sec:overview}

\para{The Stable Routing Problem}
Intuitively, a network is just a graph $G$ where nodes are routers
and edges are links between them.  The network's control plane
has a collection of router-local rules that determine how routing
messages are processed.  Different routing
protocols use different kinds of messages.  For instance,  in RIP,
a simple distance-vector protocol, messages include destination prefix
and hop count.  In contrast, 
BGP messages include a destination prefix,
an AS-path, a local preference and other
optional data. We call all such messages \emph{attributes}
regardless of their contents.
While routing protocols differ significantly in many respects,
they can be factored into two generic parts:
(1) a comparison relation that prefers certain attributes,
and (2) a transfer function that transforms incoming and outgoing messages.

An SRP instance assembles all of
these ingredients: (1) a graph defining the network topology,
(2) a destination to which to route, (3) a set of attributes, 
(4) an attribute comparison relation, and (5) an attribute transfer function.
Its \emph{solution} ($\lab$) 
associates an attribute with each node, which  
represents the route chosen by the node. Every
SRP solution has the property that nodes have not been offered 
an attribute by a neighbor that is preferred more than the chosen
one.
An SRP solution also implicitly defines a forwarding relation:  If
 $a$ receives its chosen attribute from $b$, then $a$ will
forward traffic to $b$.
There can be multiple solutions to an SRP.

\begin{figure}
  \includegraphics[width=3in]{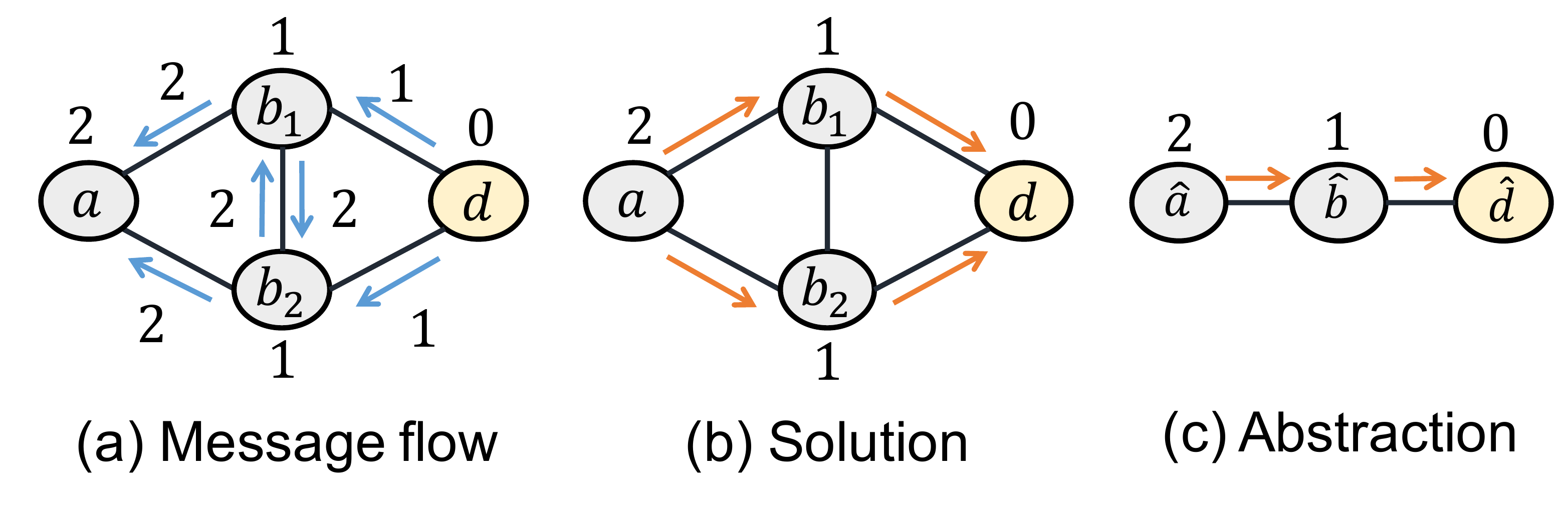}
  \vspace{-1em}
  \caption{An example RIP network. }
  \label{fig:rip}
\end{figure}

As an example, consider the RIP network in Figure~\ref{fig:rip}(a).  The destination node is  d.  It initiates
the protocol by sending messages that contain
the hop count to the destination.
The RIP comparison relation prefers the
minimal attribute (i.e., the shortest path to the destination).  The RIP
transformation function adds one to each attribute along each link. Figure~\ref{fig:rip}(b) shows the resulting solution.

\para{Network Abstractions}
Our goal is to define an algorithm that, given
one SRP, computes a new, smaller SRP that exhibits
``similar'' control plane behavior.  We call the input SRP
the \emph{concrete network}, and the output SRP the \emph{abstract
  network}.  A \emph{network abstraction} defines precisely the relationship between the 
two. 
It is a pair of functions ($f$, $h$), where $f$ is
a \emph{topology abstraction} that maps the nodes and edges of the
concrete network to those of the abstract network, and $h$ is an
\emph{attribute abstraction} that maps the concrete attributes
in control plane messages to abstract ones.

We define "similarity" using control plane solutions that emerge after
running a routing protocol. More precisely, two networks are
\emph{control-plane equivalent (CP-equivalent)} when:

\begin{quote} 
\emph{
  There is a solution $\lab$ to the concrete network \ifft there is a solution $\wlab$ to the abstract network where $(i)$ \hl{routers are labeled with similar attributes, as defined by the attribute abstraction; and} $(ii)$ \hl{packets are forwarded similarly, as defined by the topology abstraction.}
}
\end{quote}
\vspace{.3em}


\CPequivalence is  
powerful because it preserves many properties such as
reachability,  path length, way-pointing, and loop-freedom.  
Moreover, because the connection between abstract and concrete networks is tight
(\emph{i.e.}, a bisimulation) as opposed to an over-approximation, bugs found when verifying
the abstract network, correspond to real bugs  in the concrete network (\emph{i.e.}, no false positives).  Likewise,
because our abstractions are not under-approximations, if we verify that there are no violations
of a property in the abstract network, then there are no violations of the
property in the concrete network (\emph{i.e.}, no false negatives).

Figure~\ref{fig:rip}(c) shows a 
\cpequivalent abstraction
of the example network.  The topology abstraction
$f$ maps the concrete node $a$ to $\wh{a}$, $b_1$ and $b_2$ to $\wh{b}$, and $d$ to $\wh{d}$,
while the attribute abstraction $h$ is simply the identity function,
leaving hop count unchanged.  The abstraction is \cpequivalent because
there is only one stable solution to both abstract and concrete networks,
and given a concrete node $n$, the label associated with that node is the
same as the label associated with $f(n)$.  For instance, $b_1$ is labeled
with attribute 1 and so is $\wh{b}$, its corresponding node in the abstraction.
One can also observe that the forwarding relation in the concrete network is
equivalent (modulo $f$) to the forwarding relation
in the abstract network.  For instance, concrete node $b_1$ forwards to $d$
and the corresponding abstract node $\wh{b}$ forwards to $\wh{d}$ as well.


\para{Effective Abstractions}
While \cpequivalence is our goal, we cannot evaluate pairs of networks
for equivalence directly---naively, one would have to simulate the behavior of
the pair of networks on all possible inputs, an infeasible task.  
Instead, we formulate a set of conditions on network abstractions
that imply \cpequivalence and can be evaluated efficiently.
\emph{\Effective abstractions} are those that satisfy these conditions.

While these conditions help us identify abstractions for protocols such
as RIP and OSPF, there is a serious complication for BGP.
One of the conditions is \emph{transfer-equivalence},
{\em i.e.,} the routing information is transformed in a similar way in concrete and abstract networks.  
However, BGP routers employ an implicit loop-prevention mechanism
that rejects routes that contain their own AS (Autonomous System, an
identifier for the network) number.  Consequently, even when two routers
have identical configurations, their transfer functions
are slightly different because they reject different paths. 

To handle this complication, we
define a refined set of conditions,  
called the {\em \BGPeffective conditions}.
These conditions can also imply \cpequivalence and can be evaluated efficiently,
though the relationship between abstract and concrete networks is more sophisticated; the function mapping nodes in the concrete to the abstract networks is not fixed but
instead depends on the (one of possibly multiple) solutions to which the control plane converges.  

More precisely, given a concrete $SRP$ and an \effective{} abstraction, which produces $\wh{SRP}$,
a \BGPeffective{} abstraction
provides an intermediate network $\ol{SRP}$.  This intermediate network is
similar to $\wh{SRP}$ except that an abstract node $\wh{n}$ in $\wh{SRP}$  is
split into several nodes---one for each possible forwarding behavior of $\wh{n}$.  
Importantly, we prove that the number of instances node $\wh{n}$ needs to be 
split into, is bounded by $k$, where $k$ is the
number of different local preference values that the concrete nodes may use.  (Operators use local preferences to implement policy-based path selection).

\begin{figure}
  \includegraphics[width=.7\columnwidth]{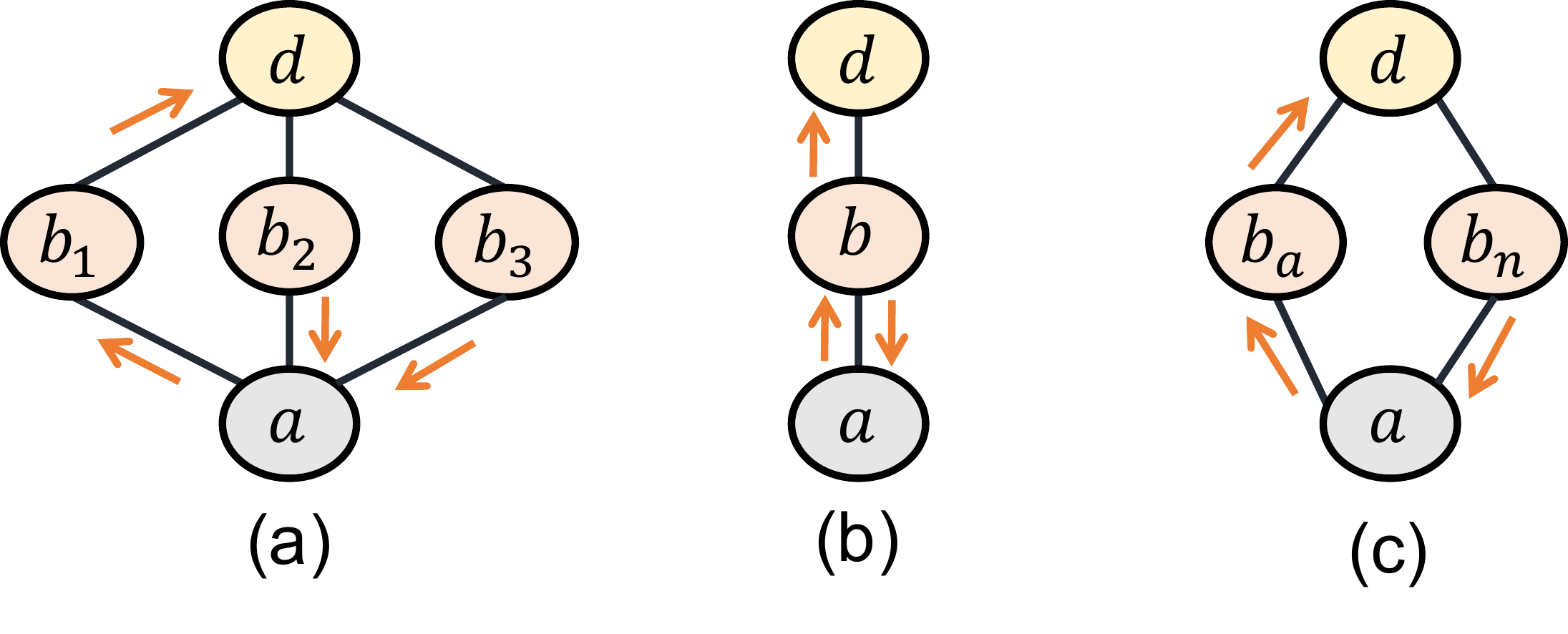}
  \vspace{-1em}
  \caption{
    (a) Concrete BGP network.
    (b) Unsound abstraction (has a loop).
    (c) Sound abstraction.
   }
  \label{fig:bgp-abstraction}
\end{figure}

Figure~\ref{fig:bgp-abstraction} shows a situation in which these
sorts of difficulties arise.  Assume the middle routers ($b_1$, $b_2$, $b_3$)
of the concrete network have identical configurations and prefer to route
traffic down rather than up.  Despite this preference, one of the three must route upwards.  In the
figure, $b_1$ happens to be that router. 
This solution is stable---no router receives a route from a neighbor
that it prefers to the current route (if router $b_1$ were to receive
a route from $a$, the path to $d$ would be $b_1 . a . b_1 . d$, a loop
which $b_1$ would reject).  And yet, despite identical configurations,
routers $b_1$ and $b_2$ forward in different directions.
Figure~\ref{fig:bgp-abstraction}(b) shows a naive (and incorrect) abstraction
in which all three of $b_1$, $b_2$ and $b_3$ are collapsed to the same
node.  This abstract network in (b) is not \cpequivalent to
the network in (a), because mapping the solution to (a) in (b)
requires generating a forwarding loop.  However, there does exist a smaller
\cpequivalent abstract network---the network depicted in Figure~\ref{fig:bgp-abstraction}(c).  The latter network is capable of mapping the solution
depicted in Figure~\ref{fig:bgp-abstraction}(a) without introducing a
forwarding loop.

\para{From Theory to Practice}
Our theory provides the basis for 
developing an efficient algorithm for control plane compression.
Based on \emph{abstraction refinement},  our algorithm first generates the coarsest possible abstraction
and then repeatedly splits abstract nodes until
the resulting network satisfies the conditions of an
(BGP-)effective abstraction.

\begin{figure}
  \includegraphics[width=\columnwidth]{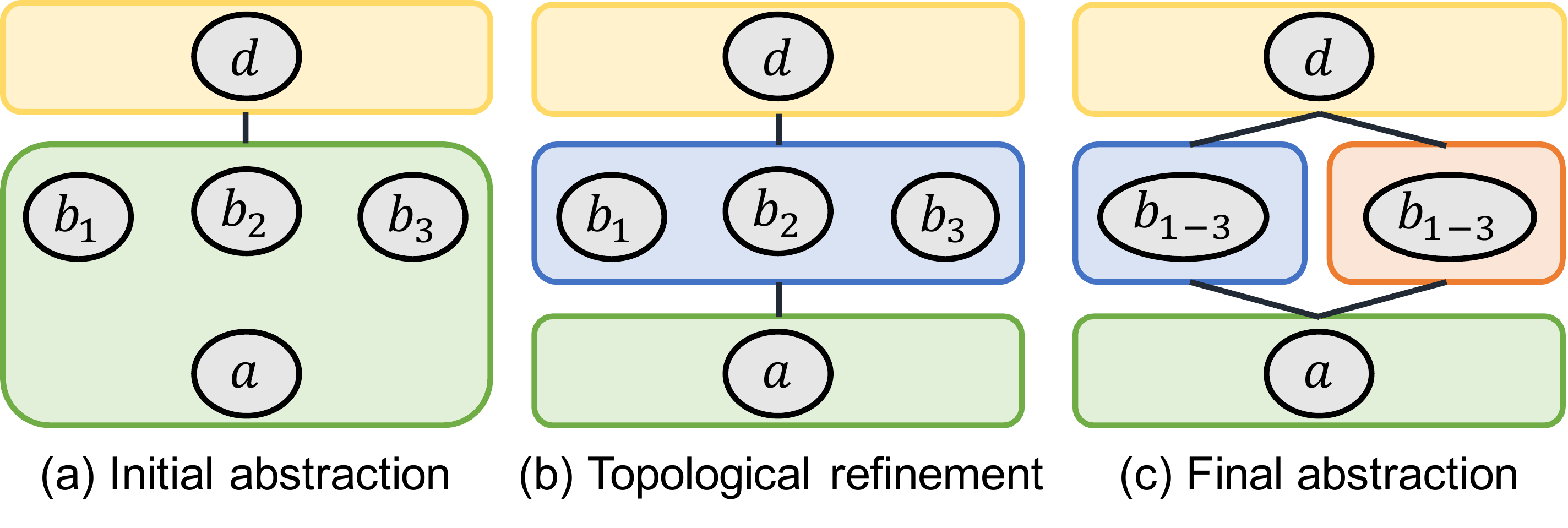}
  \vspace{-2.4em}
  \caption{Abstraction refinement for the network in Figure~\ref{fig:bgp-abstraction}(a). Boxes represent abstract nodes.}
  \label{fig:abstraction-refinement}
\end{figure}

Figure~\ref{fig:abstraction-refinement} visualizes the algorithm on the BGP network
of Figure~\ref{fig:bgp-abstraction}(a).  As a first step in Figure~\ref{fig:abstraction-refinement}(a), we generate
the coarsest possible abstraction: the destination is represented alone as
one abstract node and all other nodes are grouped in a separate abstract
node.  This first abstraction is not an effective
abstraction---it does not satisfy a topological condition
requiring that all concrete nodes ($b_1$, $b_2$, $b_3$, $a$)
associated with one abstract node have edges to some concrete node ($d$)
in an adjacent abstract node.  In this case, concrete node $a$ does not satisfy
the condition.  It is thus necessary to refine the abstraction
by separating nodes $b_1$, $b_2$, and $b_3$ from  $a$.

Figure~\ref{fig:abstraction-refinement}(b) presents the second refinement step, where the topological condition is satisfied but the
BGP-\effective conditions are not:  The nodes $b_1$, $b_2$, and $b_3$
use one non-default BGP local preference to prefer routing down rather
than up and as a consequence each node may exhibit up to two
possible behaviors.  Consequently, we must split the abstract node
for $b_1$, $b_2$, and $b_3$ into two separate nodes.  We do not know
statically the mapping of concrete to abstract nodes, so our visualization places all three concrete nodes in
each abstract node to represent all possible mappings. 

Figure~\ref{fig:abstraction-refinement}(c) happens to satisfy all conditions of a BGP-\effective abstraction.
Consequently, the refinement process terminates.  The final abstraction
includes 4 abstract nodes and 4 total edges---a reduction in size
from our concrete network with  5 nodes and 6
edges. Although this simple example does not show much reduction,
as we show later, significant reductions are possible
in larger networks.

\para{Onward}
The following sections describe our approach in 
detail. \S\ref{sec:srp} formalizes the SRP, 
\S\ref{sec:abstraction} defines \effective abstractions, and 
\S\ref{sec:algorithm} describes
the compression algorithm.
Throughout the paper there are many theorems. The proofs of these theorems can be found in the appendix.


%
%
%
%

\section{Stable Routing Problem} 
\label{sec:srp}

An SRP formally captures all the routing behaviors that a network can exhibit.  We first define it formally and then outline how it can model common routing protocols. 

\begin{figure*}[t]
\begin{footnotesize}

  \hrulefill  

  \begin{minipage}[t]{.444\linewidth}
  
  \hdr{SRP instance}{\quad \fbox{$SRP = (G, A, a_{\mathrm{d}}, \prec, \transfer)$}}
  %
  \[ \begin{array}{lclr}
    G               &=& (V,E,d) & \textit{network topology} \\
    V               & & & \textit{topology vertices} \\ 
    E               &:& V \times V & \textit{topology edges} \\
    d               &:& V & \textit{destination vertex} \\
    A               & & & \textit{routing attributes} \\
    A_{\bot}        &=& A \cup \{ \bot \} & \textit{attributes or no attribute}  \\ 
    a_{\mathrm{d}}  &:& A_{\bot} & \textit{initial route} \\
    \prec           &\subseteq& A \times A & \textit{comparison relation} \\
    \transfer       &:& E \times A_{\bot} \rightarrow A_{\bot} & \textit{transfer function} \\
  \end{array} \] 
 
  \vspace{.9em}
  \hdr{Properties of well-formed SRPs}{}  
  \[ \begin{array}{lr}
    \forall v.~ (v,v) \notin E & \textit{self-loop-free} \\

    \forall e.~ \transfer(e,\bot) = \bot & \textit{non-spontaneous} \\
  \end{array} \] 

  \vspace{1em}
  \hdr{SRP solution}{\quad \fbox{$\lab : V \rightarrow A$}}
  \[
    \lab(u) =
    \begin{cases}
      a_{\mathrm{d}} & u = d \\
      \bot & \attrs(u) = \emptyset \\ 
      a \in \attrs(u) ~\text{that is minimal by}~ (\prec), &\attrs(u) \neq \emptyset \\ 
      
    \end{cases}
  \]
  \vspace{.1em}
  \[
  \begin{array}{l@{\ }c@{\ }l}
    \attrs(u) &=& \{ a ~\vert~ (e,a) \in \choices(u) \} \\
    \choices(u) &=& \{ (e,a) ~\vert~ e=(u,v),~ a=\transfer(e,\lab(v)), a \neq \bot \} \\
    \fwd(u) &=& \{ e ~\vert~ (e,a) \in \choices(u), \hlmath{a \approx \lab(u)} \} \\ 
  \end{array} 
  \]
  \vspace{.1em} 
  \[
    \begin{array}{l}
      \hlmath{a_1 \approx a_2 \iff a_1 \not\prec a_2 \wedge a_2 \not\prec a_1} \whitespace \whitespace \whitespace \whitespace \whitespace \whitespace \whitespace \whitespace \whitespace \whitespace \whitespace \whitespace \whitespace \whitespace \whitespace \whitespace \whitespace \whitespace \whitespace \\ 
    \end{array}     
  \]

  \end{minipage}
  ~~
  \vrule
  ~~
  \begin{minipage}[t]{.55\linewidth}

  \hdr{Network abstraction}{\quad \fbox{$(f,h) : (V \rightarrow \wh{V}) \times (A \rightarrow \wh{A})$}}
  \vspace{-.2em}
  \[\begin{array}{lr}
    SRP = (G,A,a_{\mathrm{d}}, \prec, \transfer) & \textit{concrete SRP instance} \\
    \wh{SRP} = (\wh{G},\wh{A},\wh{a_{\mathrm{d}}}, \wprec, \wh{\transfer}) & \textit{abstract SRP instance} \\
    \\
    u \mapsto \wh{u} ~\equiv~ f(u) = \wh{u} & \textit{vertex abstraction notation} \\
    a \mapsto \wh{a} ~\equiv~ h(a) = \wh{a} & \textit{\attribute{} abstraction notation} \\
  \end{array} \]

  \vspace{.1em}
  \hdr{\Effective abstractions}{}
  \vspace{-.2em}
  \[
  \begin{array}{lr} 
    (d \mapsto \wh{d}) \wedge (\forall d'.~ d \neq d' \implies d' \not\mapsto \wh{d}) & 
    \textit{dest-equivalence} \\
    h(a_d) = \wh{a_d} & \textit{orig-equivalence} \\
    \forall a.~~~ h(a) = \bot \iff a = \bot & \textit{drop-equivalence} \\
    \forall a, b.~~ a \prec b \iff h(a) ~\wprec~ h(b) & 
    \textit{rank-equivalence} \\
    \forall e,a.~~ h(\transfer(e,a)) = \wh{\transfer}(f(e),h(a)) & 
    \textit{trans-equivalence} \\
    \hlmath{\forall u,v.~~ (u,v) \in E \Rightarrow (\wh{u}, \wh{v}) \in \wh{E}} & \textit{\aeabsone} \\
    \hlmath{\forall \wh{u},\wh{v}.~~ (\wh{u},\wh{v}) \in \wh{E} \Rightarrow (\forall u.~ u \mapsto \wh{u} \Rightarrow \exists v.~ v \mapsto \wh{v} \wedge (u,v) \in E)} & \textit{\aeabstwo} \\
  \end{array} 
  \]

  \vspace{.1em}
  \hdr{\BGPeffective abstractions}{}
  \vspace{-.2em}
  \[
  \begin{array}{lr} 
    \forall u,v.~~ (u,v) \in E \iff (\wh{u},\wh{v}) \in \wh{E}  & \textit{\aaabs} \\
    \forall e,a.~ e=(u,v) ~\wedge~ v \notin a.\mathrm{path} \implies & 
    \textit{transfer-approx} \\
    \whitespace \whitespace \whitespace \whitespace \whitespace \whitespace \whitespace \whitespace  \whitespace h(\transfer(e,a)) = \wh{\transfer}(f(e),h(a)) \whitespace \whitespace \whitespace  \whitespace \whitespace \whitespace \whitespace \whitespace \whitespace \whitespace  \\
  \end{array} 
  \]

  \vspace{.1em}
  \hdr{\CPequivalence}{\quad \fbox{$SRP \Similar \wh{SRP}$}}
  \vspace{-.2em}
  \[
  \begin{array}{@{}lr}
    \lab \in SRP \Longleftrightarrow \wh{\lab} \in \wh{SRP} \whitespace \mbox{when:} & \\
    \qquad 1.~~~
    \forall u.~ h(\lab(u)) = \wh{\lab}(f(u)) & 
    \textit{label-equivalence} \\
    \qquad 2.~~~ \forall u,v.~~ (u,v) \in \fwd(u) \iff (\wh{u}, \wh{v}) \in \wfwd(\wh{u}) & \textit{fwd-equivalence} \\ 
  \end{array} 
  \]

  \end{minipage}

  \hrulefill 

  \vspace{-.2em}
  \caption{Technical cheat sheet.  Definitions for SRPs, solutions, abstractions, and abstraction properties.}
  \label{fig:spp-and-abstraction}
\end{footnotesize}
\end{figure*}

\subsection{Definition and Solutions}

We define one SRP per destination in the network. 
As shown in Figure~\ref{fig:spp-and-abstraction}, an SRP instance is a tuple $(G, A, a_{\mathrm{d}}, \prec, \transfer)$. Here, $G = (V,E,d)$ is a graph with a set of vertices $V$, a set of directed edges $E : V \times V$, and a destination vertex $d \in V$.
$A$ is a set of \emph{attributes} that describe the fields of routing messages. For example, when modeling BGP, $A$ might represent tuples of a 32-bit local-preference value, a set of 16-bit community values, and a list of ASes representing the AS path. We also define a new set $A_{\bot} = A \cup \{ \bot \}$, which adds a special value $\bot$ to represent the absence of an attribute (routing message). Further, the special attribute value $a_{\mathrm{d}}$ represents the initial protocol message advertised by the destination $d$. 


In the SRP instance, $\prec$ is a partial order that compares attributes and models the routing decision procedure that compares routes using some combination of message fields.
If attribute $a_1 \prec a_2$, then $a_1$ is more desirable.
Finally, $\transfer$ represents the transfer function that describes how attributes are modified (or dropped) between routers. Given an edge and an attribute from the neighbor across the edge, it determines what new attribute is received at the current node. The transfer function depends on both the routing protocol and node's configuration. 


\para{Well-formed SRPs} 
In an SRP, the $\prec$ relation and $\transfer$ function can compare and modify attributes arbitrarily. While this generality helps model a wide variety of routing protocols, it also allows nonsensical behaviors. We define {\em well-formed} SRPs as those with two practical properties: 
(1) {\em self-loop-freedom}:  The graph must not contain self loops: $\forall v. (v,v) \notin E$. 
(2) {\em non-spontaneity}: If a neighbor has no route to the destination, then a router will not obtain a route from that neighbor. While useful, non-spontaneity is not necessary for all of our theoretical results (e.g., see SRPs for static routing).


\para{Solutions}
Given an SRP instance, we can describe its (possibly multiple) solutions.
Intuitively, each solution is derived from a set of constraints that requires that each node be \emph{locally stable}, \IE, it has no incentive to deviate from its currently chosen neighbor. For shortest path routing, an SRP solution will be a rooted tree where each node points to the neighbor with the shortest path. For policy-based routing such as BGP, the paths may not be the shortest paths.

Formally, an SRP solution is an attribute labeling $\lab : V \rightarrow A$ that maps each node to a final route (attribute) chosen to forward traffic. The labeling $\lab$ must satisfy the constraints shown in Figure~\ref{fig:spp-and-abstraction} (lower left).
The labeling of the destination node should be the special attribute $a_{\mathrm{d}}$. If there are no attributes available from neighbors ($\attrs(u) = \emptyset$), then node $u$ has no route to the destination ($\bot$). Otherwise, $\lab(u)$ is chosen to be an attribute choice that is minimal according to the comparison relation ($\prec$). If there is more than one minimal attribute, then any value can be chosen.
The set of attributes at a node stems from the choices from neighbors: for each edge $e = (u,v)$ from $u$, apply the transfer function from the neighbor's label to obtain a new attribute $a = \transfer(e,\lab(v))$, ignoring any attributes that get dropped ($a = \bot$).

Given an SRP solution, it is easy to determine the forwarding behavior. We define $\fwd(u)$ as the set of edges $e$ such that the attribute learned from $e$ is equal to the best choice $\lab(u)$ at $u$. \hl{The attribute need not be exactly} $\hlmath{\lab(u)}$, \hl{but must be at least as good} ($\approx$). If there is more than one such choice, then a node may forward to multiple neighbors.

%
%
%
%

\subsection{Modeling Common Routing Protocols}
 
 SRP can faithfully model  common routing protocols. For ease of exposition, assume for now that the network runs only one routing protocol; we consider multi-protocol networks and other configuration primitives in \S\ref{sec:practical-extensions}. 
 
   
\para{RIP (distance vector)}
RIP uses shortest paths to the destination based on hop count.
The attributes, representing the path length, are $A = \{0..15\}$ as RIP uses a maximum path length of 16; the destination attribute $a_{\mathrm{d}}$ is 0; the comparison relation  prefers shorter paths; and the transfer function drops a route if it exceeds the hop count limit and increments the path length otherwise. 

\para{OSPF (link state)}
Open Shortest Path First is a popular link state protocol where routers exchange link cost information 
and compute the least-cost path to the destination. 
The attribute set $A = \mathbb{N}$ is any natural number and represents paths cost;
the comparison relation compares this cost; and the transfer function adds the (configured) link cost. 
A large OSPF network may be split into multiple areas and prefer intra-area routes over inter-area ones. We model this behavior using attributes that are tuples of the path cost and a boolean that indicates whether it is an inter-area route. The comparison relation  prioritizes intra-area routes followed by path cost, and the transfer function changes the boolean value when crossing an inter-area edge. 


\para{BGP (path vector)}
BGP is a widely-used path-vector protocol that provides flexibility for configuring policy and computing non-shortest paths. We assume here that all routers use their own AS number, i.e., eBGP (as in large data centers~\cite{bgp-in-dc}) and discuss  iBGP in \S\ref{sec:practical-extensions}.  
We model eBGP using $A = \mathbb{N} \times 2^{\mathbb{N}} \times list(V)$, where the components are: (1) a local preference value, (2) a collection of community tags, and (3) a list of nodes defining the AS path. \hl{(Other BGP attributes such as MEDs or origin type can be modeled similarly}, but are omitted for simplicity.) BGP's comparison function first compares local-preference followed by the AS path length.  Its transfer function appends the current AS to the AS path when exporting a route.  It also drops attributes that form a loop when the current node is present in the AS path. Otherwise, the router's policy, per its configuration, is applied. 

\begin{figure}[t!]
 \includegraphics[width=.94\columnwidth]{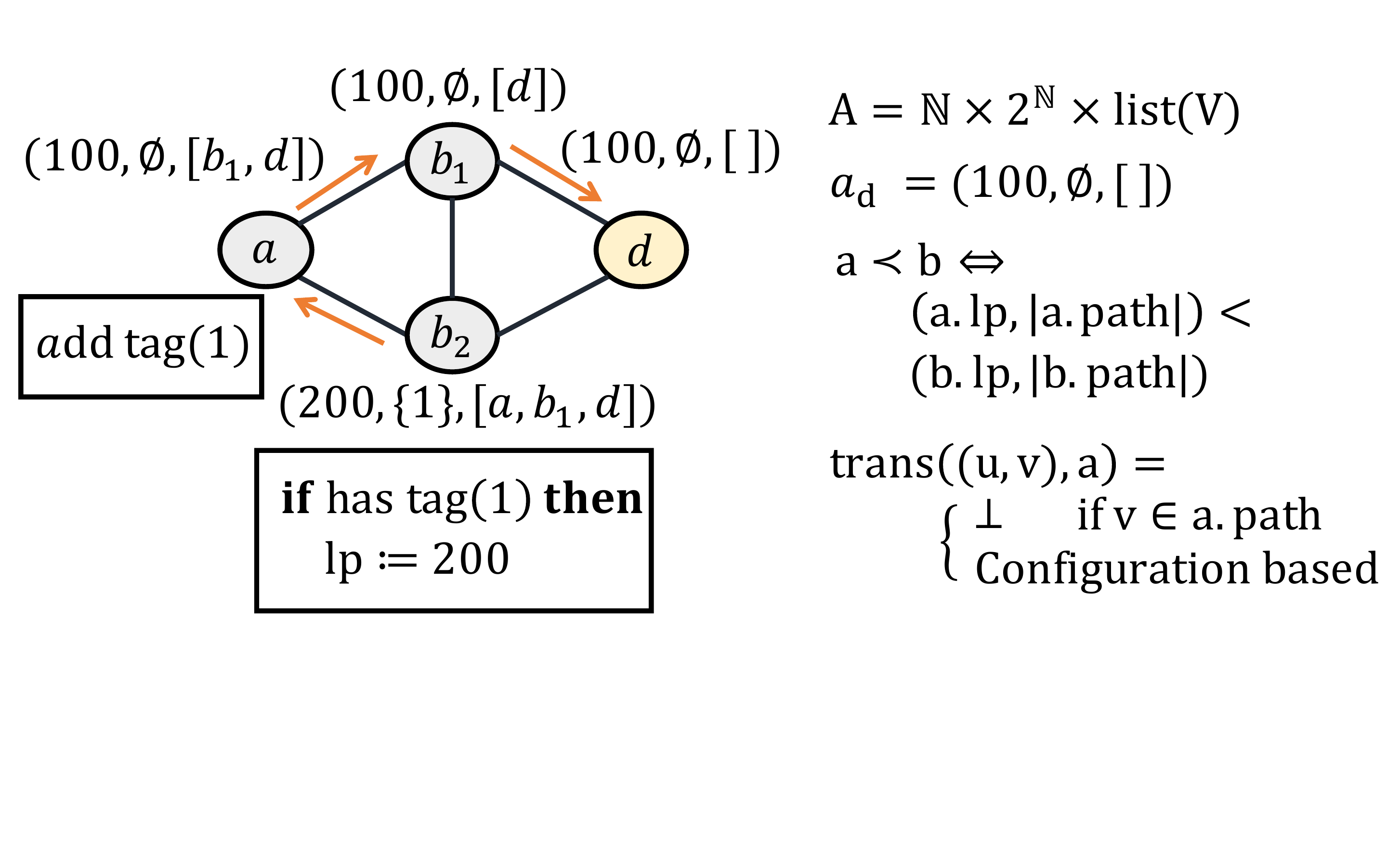}
 \vspace{-4em}
 \caption{Modeling BGP with SRP.} 
 \label{fig:srp-bgp}
\end{figure}

Figure~\ref{fig:srp-bgp} shows an example, where $a.\lp$ and $a.\aspath$ denote components of an attribute $a = (\lp,\tags,\aspath)$.  Assume that in this network $b_2$ prefers going through $a$ to reach destination $d$ and that this policy is achieved by configuring $a$ to add  tag 1 to outgoing messages and configuring $b_2$ to prefer this tag. The configuration-driven part of the transfer function is shown in the boxes for routers $a$ and $b_2$. Router $a$ adds the tag 1 to attributes it exports; and $b_2$  checks for this tag, and if present,  assigns a higher (better) local preference value than the default value (100), which ensures that $b_2$ prefers to go through $a$.  The arrows in the figure indicate the final forwarding behavior of this network, and a solution labeling $\lab$ is shown next to each node.

\begin{figure}[t!]
  \includegraphics[width=.94\columnwidth]{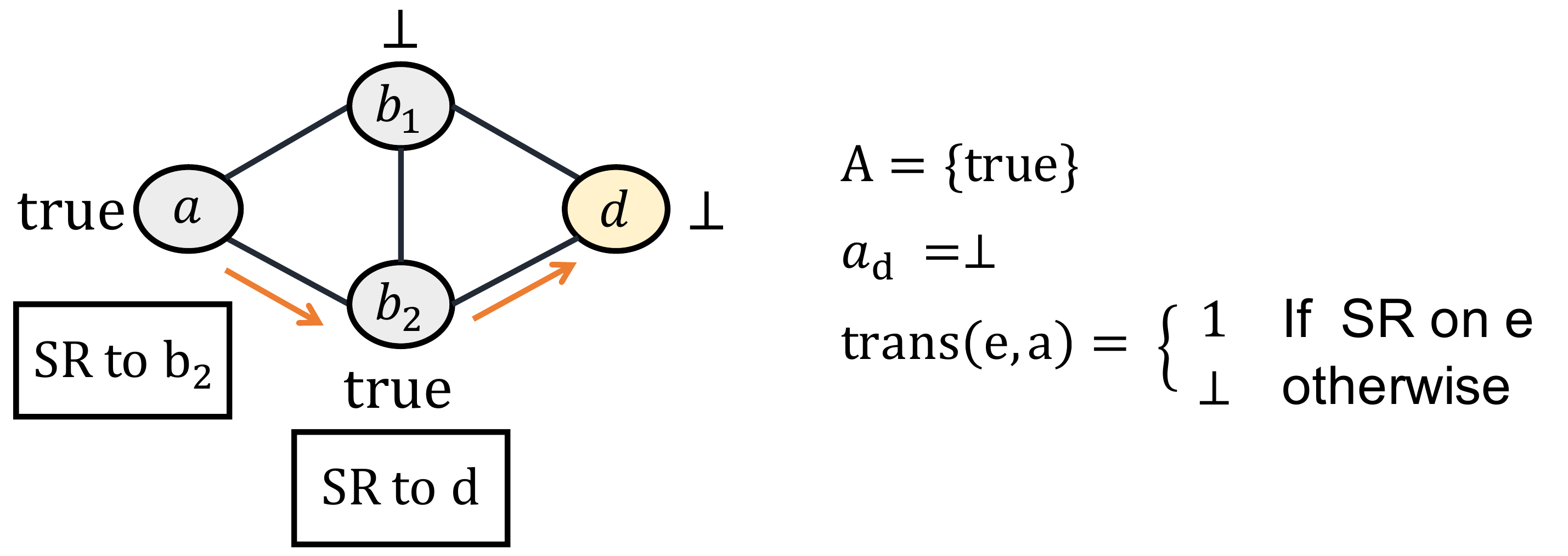}
  \vspace{-1em}
  \caption{Modeling Static routing with SRP.} 
  \label{fig:srp-static}
\end{figure}

\para{Static routing}
Operators configure static routes that describe which interface to use for a given destination. 
Figure~\ref{fig:srp-static} shows an example where routers $a$ and $b_2$ are configured with static routes. 
We model static routing using the set of attributes $A = \{true\}$ which indicates the presence of a static route. Since there is only one attribute, the comparison relation is trivially empty. The transfer function does not depend on the neighbor at all; it returns $true$ if there is a static route configured locally along an edge and $\bot$ otherwise.

%
%
%
%

\section{Effective Abstractions}
\label{sec:abstraction}



We now build on SRPs to describe the theory and implications of \effective network abstractions. 
\label{ssec:abstraction}

\para{Network abstractions} 
We start by formalizing network abstractions. A network abstraction relates two SRPs---a concrete
$SRP = (G, a_d, A, \prec, \transfer)$ and an abstract
$\wh{SRP} = (\wh{G}, \wh{a_d}, \wh{A}, \wprec, \wh{\transfer})$---using a pair of functions $(f,h)$. The topology function $f : V \rightarrow \wh{V}$ maps each concrete graph node to an abstract graph node, and the \attribute{} function $h : A \rightarrow \wh{A}$ maps each concrete attribute to an abstract one. For convenience, we will write $u \mapsto \wh{u}$ to mean $f(u) = \wh{u}$, and $a \mapsto \wh{a}$ to mean $h(a) = \wh{a}$.  We also freely apply $f$ to edges and paths:  given an edge $e=(u,v)$, $f(e)$ means $(f(u),f(v))$; given a path $u_1, \ldots, u_n$,  $f(u_1, \ldots, u_n)$ means $f(u_1), \ldots, f(u_n)$.

\begin{figure}
  \includegraphics[scale=.3]{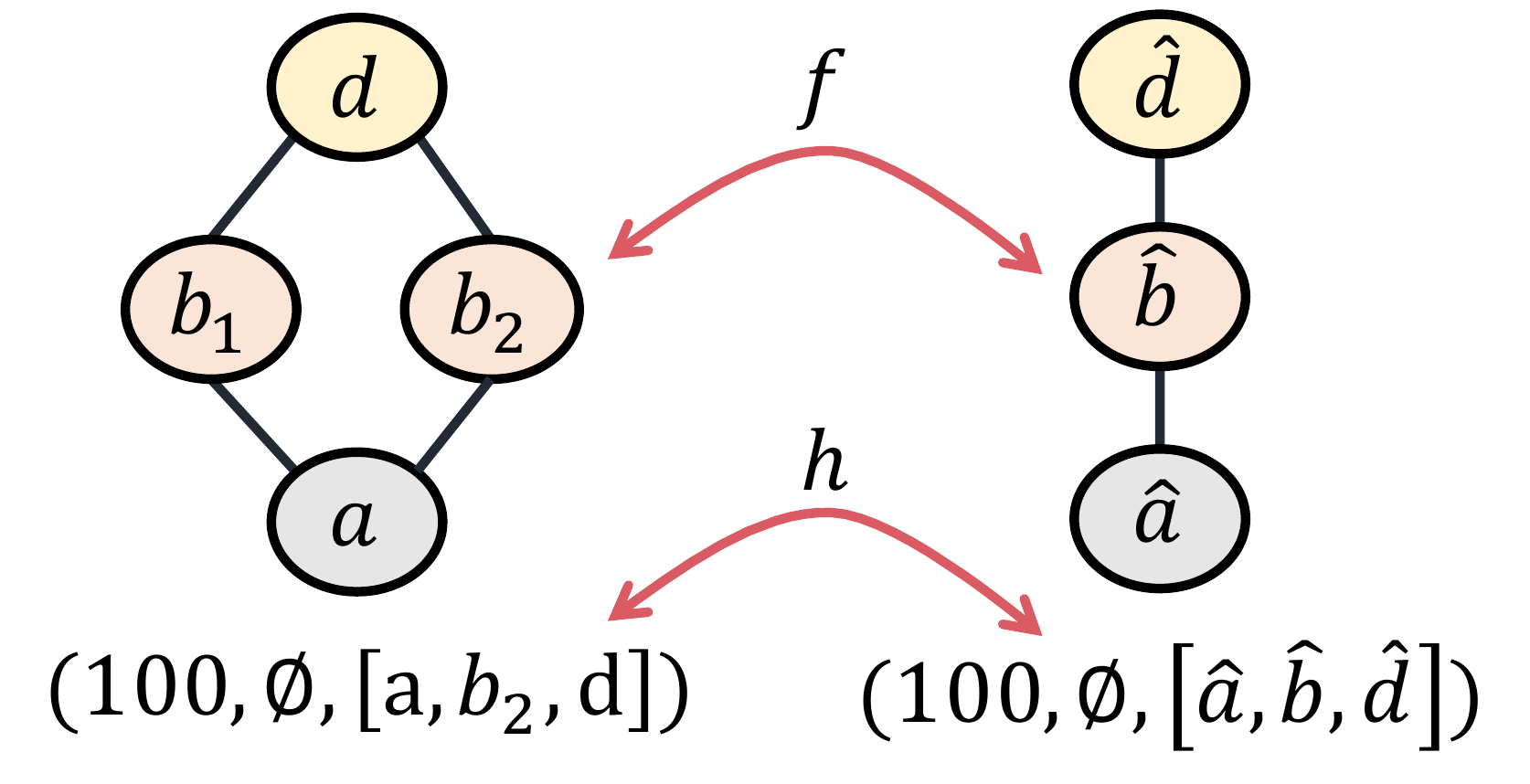}
  \vspace{-1em}
  \caption{Example abstraction for a BGP network.}
  \label{fig:spp-hfunction}
\end{figure}

\Attribute{}  abstraction allows the set of attributes to differ
between the concrete and abstract networks. This ability
may be used to convert attributes with concrete nodes
into those with related abstract nodes.  For example, in the BGP network in
Figure~\ref{fig:spp-hfunction}, $f$ maps $b_i$ nodes
to the abstract node $\wh{b}$, while $h$ maps the concrete AS path to its abstract counterpart. 

\subsection{\Effective{} Abstraction Conditions}
\label{ssec:effective}

In a network abstraction, $f$ and $h$ can be arbitrary
functions, but we are interested only in abstractions that preserve the control plane
behavior of the concrete network.  An \emph{\effective{}} abstraction 
satisfies a set of relatively easy-to-check conditions that imply \cpequivalence. 
These conditions, listed in the middle right of Figure~\ref{fig:spp-and-abstraction}, 
are restrictions on the topology function $f$ and the \attribute{}
function $h$.

\begin{figure}
  \includegraphics[scale=.3]{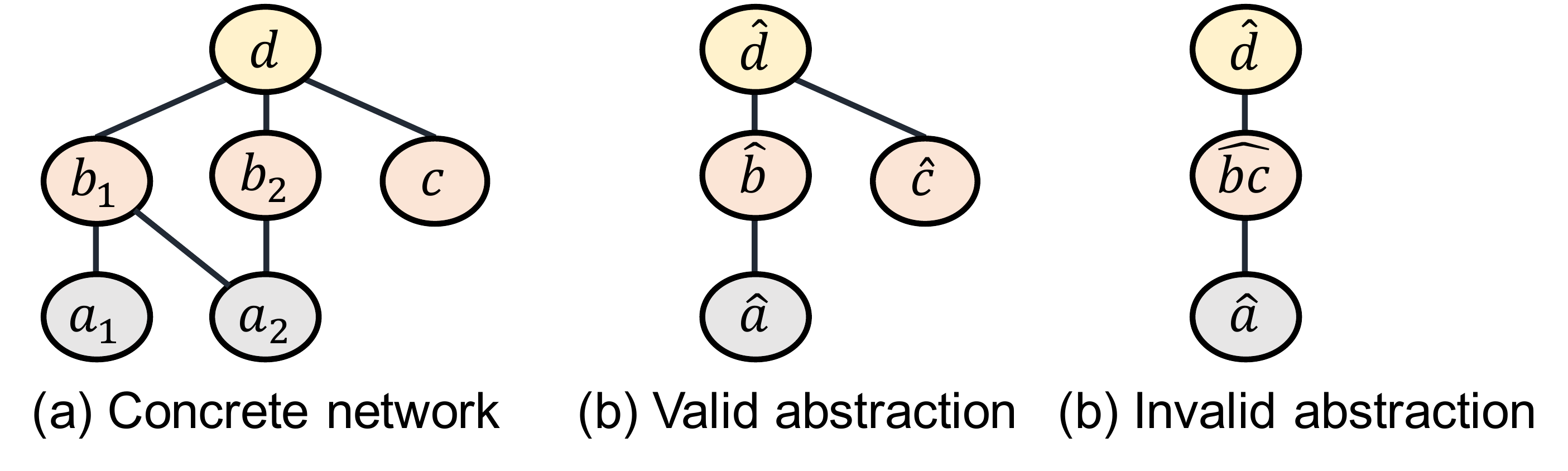}
  \vspace{-2.5em}
  \caption{Topology abstraction.}
  \label{fig:topology-abstraction}
\end{figure}

\para{Topology abstraction conditions}  
\Effective{} topology functions obey two conditions.  First, they
preserve the identity of the destination node (\emph{dest-equivalence}).
That is, the concrete destination node, and only this node, should be mapped to the abstract destination: $d \mapsto \wh{d}$,
$d' \not\mapsto \wh{d}$.
Second, the topological mapping as a whole must be a (forall-exists) $\aeabs$. A $\aeabs$ (\hl{both} \hlmath{\aeabsone} \hl{and} \hlmath{\aeabstwo})
demands that: (1) for every concrete edge $(u,v)$ there is a corresponding abstract edge $(\wh{u}, \wh{v})$ and (2) for every abstract edge ($\wh{u}$, $\wh{v}$), \emph{all} concrete nodes $u$ where $u \mapsto \wh{u}$ must have an edge to \emph{some} concrete node $v$ where $v \mapsto \wh{v}$. 
Figure~\ref{fig:topology-abstraction} shows an example of both a valid and invalid \aeabs. The abstraction on the right is invalid because $c$ does not have an edge to either $a_1$ or $a_2$ despite there being an edge between $\wh{bc}$ and $\wh{a}$ in the abstract network.
%
%
%

\para{\Attribute{} abstraction conditions}
The first conditions for \attribute{} abstraction, \emph{drop-equivalence} and \emph{orig-equivalence}, state that the abstraction function must preserve the ``no route'' and
the destination attributes: $h(\bot) = \bot$ and $h(a_d) = \wh{a_d}$.
An abstraction must also preserve the comparison relation's \attribute{} ordering
(\emph{rank-equivalence}). Finally, an abstraction must preserve the transfer
function (\emph{transfer-equivalence}), that is, applying the concrete transfer function 
and abstracting the resulting attribute should be the same as abstracting the attribute first, 
and then applying the abstract transfer function.
A critical aspect here is that, unlike \cpequivalence,
which is a network-wide property, transfer-equivalence is a simple, local property
that can be evaluated efficiently by comparing the transfer functions.  


\subsection{Effectiveness implies CP Equivalence}
\label{ssec:fwd-equivalent}

We are now ready to prove that \effective abstractions guarantee \cpequivalence in two steps. 
First, we demonstrate that \effective{} abstractions are \emph{label-equivalent}
(Figure~\ref{fig:spp-and-abstraction}).
In other words, for each solution $\lab$ to $SRP$, there exists a
corresponding solution $\wlab$ to the abstract $\wh{SRP}$
(\emph{i.e.}, whenever $\lab$ labels $u$ with $a$, $\wlab$ labels $f(u)$ with
$h(a)$), and vice-versa.  Next, we show that given related labellings, the final control plane behaviors are also related, \IE, they are equivalent with respect to forwarding (\emph{fwd-equivalent} as defined in
Figure~\ref{fig:spp-and-abstraction}). 


Our proof depends on the structure of the SRPs and
their solutions.  In particular, when the SRP nodes dynamically transmit information to
one another, we would like to be able to carry out the proof using a induction.  
However, we cannot do that if the SRP solutions contain loops, as the
induction would not be well-founded.  Fortunately,
most broadly-used dynamic routing protocols
are loop-free by design.
We will consider the simpler case of
static routes, which can be configured to create loops, separately.

{\iffull

\begin{defn}
We say that an abstraction $(f,h)$ is \emph{choice-equivalent} if the following property holds:
$$\forall v,e.~ \big( \exists a.~ a \mapsto A ~\wedge~ (e,a) \in \choices(v)$$
$$\iff$$
$$(f(e),A) \in \wh{\choices}(f(v)) \big)$$
\end{defn}

\begin{theorem} \label{thm:pref-equiv}
If we have a self-loop-free $SRP$ and $\wh{SRP}$, and a \effective abstraction that is choice-equivalent, then the abstraction is label-equivalent.

\begin{proof}
  Looking at the definition of $\lab$, there are 3 cases to consider. First we observe that if $v = d$, then $\lab(d) = a_{d}$. It follows that $\wh{\lab}(f(d)) = \wh{\lab}(\wh{d}) = \wh{a}_{d} = h(a_{d}) = h(\lab(d))$. In the second case, using choice-equivalence, we can see that $\attrs(v) = \emptyset \iff \wh{\attrs}(f(v)) = \emptyset$. Thus, $h(\lab(v)) = h(\bot) = \bot = \wh{\lab}(f(v))$. For the final case, we show the implications separately.
  
  \vspace{.8em}
  \noindent \textbf{Case ($\Rightarrow$)}
  Assume $\lab(v) = a$. By the definition of $\lab$, we know that $a \in \attrs(v)$ and is minimal by $\prec$. From choice-equivalence and rank-equivalence, we know that $A = h(a) \in \wh{\attrs}(f(v))$ that is minimal by $\wprec$. By the definition of $\lab$, we then know that $\wh{\lab}(f(v)) = A = h(a)$. By transitivity, $\wh{\lab}(f(v)) = h(\lab(v))$

  \vspace{.8em}
  \noindent \textbf{Case ($\Leftarrow$)}
  Assume $\wh{\lab}(f(v)) = A$. We know that $A \in \wh{\attrs}(f(v))$ and is minimal by $\wprec$. Assume $(f(e),A) \in \wh{choices}(f(v))$. From choice-equivalence, we know that the exists an $a$ such that $h(a) = A$ and $(e,a) \in choices(v)$. Therefore, $a \in \attrs(v)$. Let us consider a smallest such $a$ in terms of $\prec$. From rank-equivalence, we know that $a$ is smaller than any $(e',a')$ where $f(e') \neq f(e)$ since $A$ was the smallest such value in $\wh{choices}$. Therefore, $a \in attrs(v)$ and is minimal by $\prec$. Therefore, $\lab(v) = a$. It follows from transitivity that $h(\lab(v)) = \wh{\lab}(f(v))$.
\end{proof}
\end{theorem}

\begin{theorem} \label{thm:forward-equiv}
If we have a self-loop-free $SRP$ and $\wh{SRP}$ and a \effective abstraction that is choice-equivalent, then the abstraction is fwd-equivalent.

\begin{proof} \text{ }

  From Theorem~\ref{thm:pref-equiv}, we know that we have label-equivalence. 
  
  \vspace{.8em}
  \noindent \textbf{Case ($\Rightarrow$)}
  Assume $e = (u,v) \in \fwd(u)$. By the definition of $\fwd$, this means $\exists a. (e,a) \in \choices(u) \wedge \lab(u) = a$. From choice-equivalence, this means that 
  $$\exists a. (f(e),h(a)) \in \wh{\choices}(f(u)) \wedge \lab(u) = a$$
  Rewriting slightly, this would be: 
  $$\exists a. (f(e),h(\lab(u))) \in \wh{\choices}(f(u))$$ 
  By label-equivalence, we can rewrite this formula as: 
  $$(f(e),\wh{\lab}(f(u))) \in \wh{\choices}(f(u))$$
  By the definition of $\wh{\fwd}$, this is the same as:
  $$f(e) \in \wh{\fwd}(f(u))$$

  \vspace{.8em}
  \noindent \textbf{Case ($\Leftarrow$)}
  Assume $f(e) \in \wh{\fwd}(f(u))$. By the definition of $\fwd$, this means: $\exists A.~ (f(e),A) \in \wh{\choices}(f(u)) \wedge \wh{\lab}(f(u)) = A$.
  From choice-equivalence, this means:
  $$\exists a.~ h(a) = A \wedge (e,a) \in \choices(u) \wedge \wh{\lab}(f(u)) = A$$
  Rewriting slightly, we get:
  $$\exists a.~ (e,a) \in \choices(u) \wedge \wh{\lab}(f(u)) = h(a)$$
  Applying preference-equivalence, we get:
  $$\exists a.~ (e,a) \in \choices(u) \wedge h(\lab(u)) = h(a)$$

  There must exist some $(e,b) \in \choices(u)$ where $\lab(u) = b$. 
  Suppose this is not the case. Then there must be some $(e',b') \in \choices(u)$ with $\lab(u) = b'$ and $f(e') \neq f(e)$. However, this would mean that $h(\lab(u)) = h(b') = \wh{\lab}(f(u))$. From rank-equivalence, this would mean that $h(b') \wprec h(a)$, which contradicts the fact that $\wh{\lab}(f(u)) = h(a)$.
  Therefore, it must be the case that there is some $(e,b) \in \choices(u)$ with $\lab(u) = b$. 
  By definition this means that $e \in \fwd(u)$.
  This satisfies fwd-equivalence. 
\end{proof}
\end{theorem}
\fi}



\begin{theorem} \label{thm:dag}
  Any solution $\lab$  to a well-formed, loop-free SRP will form a DAG rooted at the destination $d$.
  {\iffull
  \begin{proof}
    We know that the solution is loop-free so the result must not have cycles. Also, there can only be one root for the DAG ($d$) because if there were another $d'$, then $\lab(d') = \bot$, otherwise $d'$ would forward to some neighbor. However, because the SRP is non-spontaneous, this can not happen.
  \end{proof}
  \fi}
\end{theorem}

Using this property of stable solutions, we can prove that for any
concrete solution $\lab$, there is an abstract solution $\wlab$ such that the solutions
are label- and fwd-equivalent (and vice-versa). The proof goes in two steps. First, we
prune the graph to include only edges in $\lab$ or $\wlab$ that are involved in
forwarding.  Within such subgraphs, we can show by induction on the length of the
forwarding paths that the subgraphs satisfy label-equivalence and fwd-equivalence.
It is then easy to come to our desired conclusion by
showing that adding the removed edges back in does not affect the stable solution
of either the concrete or the abstract graph.

\begin{theorem} \label{thm:correctness-local}
  A well-formed, loop-free $SRP$ and its \effective abstraction $\wh{SRP}$
  are label- and fwd-equivalent.

  {\iffull
  \begin{proof} 
    It suffices to first show choice-equivalence. We then get label-equivalence for free from Theorem~\ref{thm:pref-equiv}, and then that $SRP$ and $\wh{SRP}$ are fwd-equivalent from Theorem~\ref{thm:forward-equiv}. Thus, we need to show:
    $$\exists a.~ h(a) = A \wedge (e,a) \in \choices(v) \iff (f(e),A) \in \wh{\choices}(f(v))$$

    Because we know the SRP is loop-free and non-spontaneous, we know that any stable solution $\lab$ (and $\wlab$) must form a rooted DAG at the destination $d$ (Theorem~\ref{thm:dag}). 
    
    We start by showing a slightly strengthened inductive hypothesis: with the choice-equivalence property above for all $e = (u,v)$ where $e$ goes from level $k+1$ to level $k$ in the DAG, while simultaneously showing that label-equivalence ($h(\lab(v)) = \wh{\lab}(f(v))$) holds at each node. We show each direction of the stronger implication separately, using induction on the level of the DAG.
    
    \vspace{.8em}
    \noindent \textbf{Base case (for $\Rightarrow$ and $\Leftarrow$):}
    For the base case, from the definition of $\lab$, we know that $\lab(d) = a_{d}$ and $\wh{\lab}(\wh{d}) = \wh{a}_{d}$. 
    From dest-equivalence, we know that $f(d) = \wh{d}$, so:
    $$\wh{\lab}(f(d)) = \wh{\lab}(\wh{d}) = \wh{a}_{d} = h(a_{d}) = h(\lab(d))$$
    
    Since there are no edges $e$ going to a lower level in the DAG (than the root) in either the concrete or abstract, we are done.
    
    \vspace{.8em}
    \noindent \textbf{Inductive case ($\Rightarrow$)}
    Suppose $(e,a) \in \choices(v)$ and $e = (u,v)$. We know that $a = \transfer(e)(\lab(v)) \neq \bot$. By the IH with preference-equivalence, we know that $\wh{\lab}(f(v)) = h(\lab(v))$. From transfer-equivalence, we know that 
    $$\wh{\transfer}(f(e))(h(\lab(v))) = h(\transfer(e)(\lab(v))) = h(a) \neq \bot$$
    By transitivity, we know:
    $$\wh{\transfer}(f(e))(\wh{\lab}(f(v))) = h(a)$$
    By the definition of $\choices$, it follows that
    $$(f(e),h(a)) \in \wh{\choices}(f(v))$$
    Since we have choice-equivalence, we have fwd-equivalence for such edges $e$ going down in the DAG.

    \vspace{.8em}
    \noindent \textbf{Inductive case ($\Leftarrow$)}
    Suppose $(E,A) \in \wh{\choices}(f(v))$.
    From the \aeabs, we know there is some $e$ such that $f(e) = E = (f(u),f(v))$. We also know that $A = \wh{\transfer}(f(e))(\wh{\lab}(f(v))$
    As before, we observe by the IH that $\wh{\lab}(f(v)) = h(\lab(v))$
    And so:
    $$A = \wh{\transfer}(f(e))(h(\lab(v))) = h(\transfer(e)(\lab(v)))$$
    From choice-equivalence, we know there exists some value $a$ where $a = \transfer(e)(\lab(v))$
    By the definition of $\choices$, it follows that:
    $$\exists a.~ h(a) = A \wedge (e,a) \in \choices(v)$$
    Since we have choice-equivalence, fwd-equivalence.

    \vspace{.8em}
    \noindent \textbf{Other edges}
    All that remains is to show that edges going to a equal or higher level of the DAG also satisfy choice-equivalence. Because we have shown label-equivalence for edges going down in the DAG, and due to the definition of forwarding ($\mathrm{fwd}$), edges not decreasing in the depth of the DAG can not have been chosen as the best route. This means that we have label-equivalence even when now considering these edges as well. Because the labels do not change, choice-equivalence then follows trivially from the definition of $\choices$ and label-equivalence -- the reasoning is the same as for the inductive case.
  \end{proof}
  \fi}
\end{theorem}

Using Theorem~\ref{thm:correctness-local}, we may also conclude that
any \effective{} abstractions of common protocols, which produce loop-free routing, are
\cpequivalent.  
However, effectiveness requires transfer-equivalence, which as mentioned previously commonly does not hold for BGP. That makes it impossible to obtain effective abstractions for BGP networks. 
In the next subsection, we address this 
shortcoming by defining another kind of abstraction that is applicable for BGP. 

  {\iffull
  \begin{proof}
    Because models of each of these routing protocols define well-formed and loop-free
    SRPs, by Theorem~\ref{thm:correctness-local},
    their \effective{} abstractions are fwd-equivalent.
  \end{proof}
  \fi}



\para{Static routing}
Networks with
static routes are not necessarily loop-free. (The presence of a loop would clearly
be a bug, but we must be sure our theory is sound in such a situation so we
can use it to detect inadvertent bugs caused by misconfiguration of static routes.)
Fortunately, due to the simple nature of static routing---static routes do not depend on other routes learned from neighbors---we can prove its correctness independently.

\begin{theorem} \label{thm:correctness-static}
  A self-loop-free $SRP$ and $\wh{SRP}$ for static routing with an
  \effective{} abstraction will have fwd-equivalence.
  
  {\iffull
  \begin{proof}
    Because the labeling at each node does not depend on the labeling at other nodes,
    the proof is direct. 
    As before, we show choice-equivalence, then rely on~\ref{thm:forward-equiv} to derive \cpequivalence.

    \vspace{.8em}
    \noindent \textbf{Case ($\Rightarrow$)}
    Assume $e = (u,v)$. We have $(e,a) \in \choices(u)$ and $h(a) = A$. By unfolding the definition of $\choices$, we know that $a = \transfer(v)(\lab(v))$. By transfer equivalence, we know that 
    $$h(a) = h(\transfer(e)(\lab(v))) = \wh{\transfer}(f(e))(h(\lab(v)))$$
    There are now 2 cases. Suppose $a = 1$. Then $h(a) = 1$, so 
    $$\wh{\transfer}(f(e))(h(\lab(v))) = 1$$
    Since the definition of $\transfer$ does not depend on the attribute for static routes, we know that :
    $$\wh{\transfer}(f(e))(\wh{\lab}(v)) = 1$$
    It follows that 
    $(f(e),1) \in \choices(f(u))$
    
    \noindent The case for $a = 0$, is symmetric.

    \vspace{.8em}
    \noindent \textbf{Case ($\Leftarrow$)}
    Suppose $(f(e), A) \in \choices(f(u))$. We need to show that there exists an $a$ such that $(e,a) \in \choices(u)$ and $h(a) = A$. Let us choose $A = a$ for static routes. Clearly $h(a) = A$ since $h$ is the identity. We have:
    $$A = h(a) = a = \wh{\transfer}(f(e))(\wh{\lab}(f(v))))$$
    Again, since the definition of $\wh{\transfer}$ does not depend on the attribute, this is the same as:
    $$\wh{\transfer}(f(e))(h(\lab(v)))$$
    From transfer-equivalence and transitivity we know that:
    $$a = \transfer(e)(\lab(v))$$
    Finally, from the definition of $\choices$:
    $\exists a.~ h(a) = A \wedge (e,a) \in \choices(u)$
  \end{proof}
  \fi}
\end{theorem}

\OMIT{
Finally, we obtain a general property that lifts
\cpequivalence to properties of forwarding paths.

\begin{corollary} \label{thm:all-protocol-equiv}
  Suppose we have a self-loop-free $SRP$ and $\wh{SRP}$ for RIP, OSPF,  static routing, or BGP, related by
  \effective{} abstraction $(f,h)$.
  There is a solution $\lab$, where each node $u_1 \mapsto \wh{u_1}$ forwards along label path $s = \lab(u_1) \ldots \lab(u_k)$ to some node $u_k \mapsto \wh{u_k}$ \ifft there is a solution $\wlab$ that forwards along the label path $\wh{s} =  \lab(\wh{u_1}) \ldots \lab(\wh{u_k})$ and $h(s) = \wh{s}$.

  {\iffull
  \begin{proof} We show each direction separately.

    \vspace{.8em}
    \noindent \textbf{Case ($\Rightarrow$)}
    Suppose $\lab$ is a solution for $SRP$.      Given any two nodes $u$ and $v$ where $u$ can reach $v$, there exists a path $p = u,w_1,\ldots,w_k,v$ where $(u,w_1) \in \fwd(u)$ and $(w_i,w_{i+1}) \in \fwd(w_i)$ and $(w_k,v) \in \fwd(w_k)$. Because $\lab$ and $\wh{\lab}$ are fwd-equivalent, we know that $(f(u),f(w_1)) \in \wfwd(f(u))$ and so on. 
    Therefore, there is an abstract path in $\wh{\lab}$ where $f(u)$ can reach $f(v)$ where the path has the form $f(u), f(w_1), \ldots, f(w_k), f(v)$. 
    The labels of the concrete path are $s = \lab(u), \lab(w_1), \ldots, \lab(w_k), \lab(v)$. Similarly, the abstract path has labels $\wlab(f(u)), \ldots, \wlab(f(v))$. It follows from label-equivalence that 
    $\wlab(f(u)), \ldots, \wlab(f(v)) = h(\lab(u)), \ldots, h(\lab(v))$. Finally, the definition of $h$ gives us: 
    $h(\lab(u)), \ldots, h(\lab(v)) = h(s)$

    \vspace{.8em}
    \noindent \textbf{Case ($\Leftarrow$)}
    Symmetric to the first case.

    Suppose $\wh{L}$ is a solution for $\wh{SRP}$. Consider an arbitrary path $\wh{u},\wh{w_1},\ldots,\wh{w_k},\wh{v}$. Then we know $(\wh{u},\wh{w_1}) \in \wh{\fwd}(\wh{u})$ and so on. 
    From the fact that $\lab$ and $\wh{\lab}$ are fwd-equivalent, every node $u$ where $u \mapsto \wh{u}$ will follow some path $(u,w_1) \in fwd(u)$ and so on. Therefore, there will be a concrete path $u, w_1, \ldots, w_k, v$ such that $w_i \mapsto \wh{w_i}$, and $v \mapsto \wh{v}$. The abstract path $\wh{s} = \wlab(\wh{u}), \ldots, \wlab(\wh{v})$. Similarly, the concrete path will have $s = \lab(u), \ldots, \lab(v)$. To show that $\wh{s} = h(s)$, we simply use label-equivalence:
    $h(s) = h(\lab(u)), \ldots, h(\lab(v)) = \wh{s}$.

  \end{proof}
  \fi}

\end{corollary}

The corollary relates paths of node \emph{labels}.
This is done so that we can relate properties both of the data plane (forwarding) as well as the control plane (labels).\footnote{It is always possible to include the neighbor through which a route is learned in the attribute itself (i.e., by adding a next-hop attribute field).}
}

\subsection{BGP with Loop Prevention}
\label{ssec:bgp-loop-prevention}

We model BGP using an abstraction: 
$h((\lp,\tags,\aspath)) = (\lp,\tags,f(\aspath))$.
%
BGP's loop-prevention is problematic here because it depends on the actual concrete path used, which implies that two concrete nodes $x$ and $y$ with
syntactically identical configurations will actually have different transfer
functions and violate transfer-equivalence.  Node $x$ will reject paths that have gone through $x$ but not $y$, and node $y$ will reject paths that have gone through $y$ but not $x$.  If we were somehow able to abstract away loop prevention, we could attempt to have topology abstractions for BGP that are transfer-equivalent. This observation motivates the additional properties laid out for BGP in Figure~\ref{fig:spp-and-abstraction}.

\para{\BGPeffective{} abstractions} For BGP, we require dest-,
drop-, orig- and rank-equivalence as for
ordinary \effective{} abstractions.  However, as opposed to a \aeabs,
we require a slightly
stronger (forall-forall) \aaabs. This constraint requires
that whenever there is an abstract edge between $\wh{u}$ and
$\wh{v}$ if and only if there is a concrete edge between $u$ and $v$.
This strong condition on the network topology allows us to get away with
a weaker condition than transfer-equivalence:  we relax
the transfer-equivalence condition to what we call transfer-approx.
The latter condition is similar to transfer equivalence,
except it ignores differences caused by BGP loop-prevention.  
Formally,
it is specified as:
\[
\begin{array}{l}
\hspace{-.4em}
\forall e,a.~ e=(u,v) \wedge v \notin a.\mathrm{path} \Rightarrow \\
\whitespace \whitespace h(\transfer(e,a)) = \wh{\transfer}(f(e),h(a))
\end{array}
\]

\para{Bounded behaviors}
Now, given a \BGPeffective{} abstraction,
we know that, when loop-prevention happens, there may be differences between the forwarding
behaviors of different concrete nodes even when they have identical configurations.  
Fortunately, we can bound the number of different behaviors that can arise dynamically,
and, moreover, we can infer that bound directly from the configurations. 
%

First, let $\behaviors_\lab(\wh{u})$ be the set of possible behaviors of concrete
nodes related to abstract node $\wh{u}$.  Second,
let $\prefs(v)$ be the set of BGP local-preference values
that may be assigned to an announcement at node $v$.
For example, if a configuration explicitly sets the local-preference
value to $200$ or $300$ depending on the route, and 100 is the default local preference,
then the set $\prefs(v) = \{ 100, 200, 300 \}$.
With these definitions in hand, we can prove the following theorem.




\begin{theorem} \label{thm:bounded-behaviors}
  If a well-formed $SRP$ and $\wh{SRP}$ for BGP has an \aaabs and is transfer-approx, then
  for all solutions $\lab$ to $SRP$, and all abstract nodes $\wh{u} \in \wh{V}$, $\abs{\behaviors_\lab(\wh{u})} \leq \abs{\prefs(\wh{u})}$.

{\iffull
\begin{proof}
  Because we have rank-equivalence and an \aaabs, the only way two nodes will forward to different neighbors is the transfer functions are different. Otherwise, both nodes would receive the same $\choices$ as in Theorem~\ref{thm:correctness-local} and because of the universal abstraction, they both have an edge to the best such choice and will use this neighbor. 
  Due to relative-transfer-equivalence, the only time this can occur is when two nodes have different transfer functions due to loop prevention.
  
  First we show that there can be $|\prefs(\wh{u})|$ different behaviors. Consider the example in Figure~\ref{fig:bounded-behaviors}. In the example, $\wh{u}$ has a local preference for $\wh{v}_1$ over $\wh{v}_2$ over $\wh{v}_3$ etc. In this case, $|\prefs(\wh{u})| = 3$. There is a stable solution where $u_1$ forwards to $v_{11}$ since that is the best path. $u_2$ would prefer to use this path, but cannot because it is already on the path, so it cannot consider $v_{11}$ due to its transfer function. Instead, $u_2$ will use the next best choice $v_{21}$. Similarly, $u_3$ would like to use $v_{11}$ or $v_{21}$ but cannot due to loops. Therefore, $u_3$ will forward to $v_{31}$ instead. 

  Because there is a universal abstraction (full mesh), and because we have rank-equivalence and relative-transfer-equivalence, each node has the same choices modulo loops. Such a chain as shown in Figure~\ref{fig:bounded-behaviors} is the only way we can get such different behavior.

  Now we show that there can not be more than $|\prefs(\wh{u})|$ behaviors. The proof is by contradiction. Suppose we have another node $u_4$ and $u_4$ will forward to a different node that each of $u_1$ through $u_3$. $u_4$ can not continue the chain by falling back to the next lowest local preference since all local preferences have been exhausted by $u_1$ through $u_3$. Therefore, $u_4$ will forward to one of the same neighbors as $u_1$ through $u_3$. But this contradicts the assumption.
  Therefore, there can not be more than $|\prefs(\wh{u})|$ behaviors.


   
\end{proof}
\fi}

\end{theorem}

\begin{figure}
  \includegraphics[scale=.31]{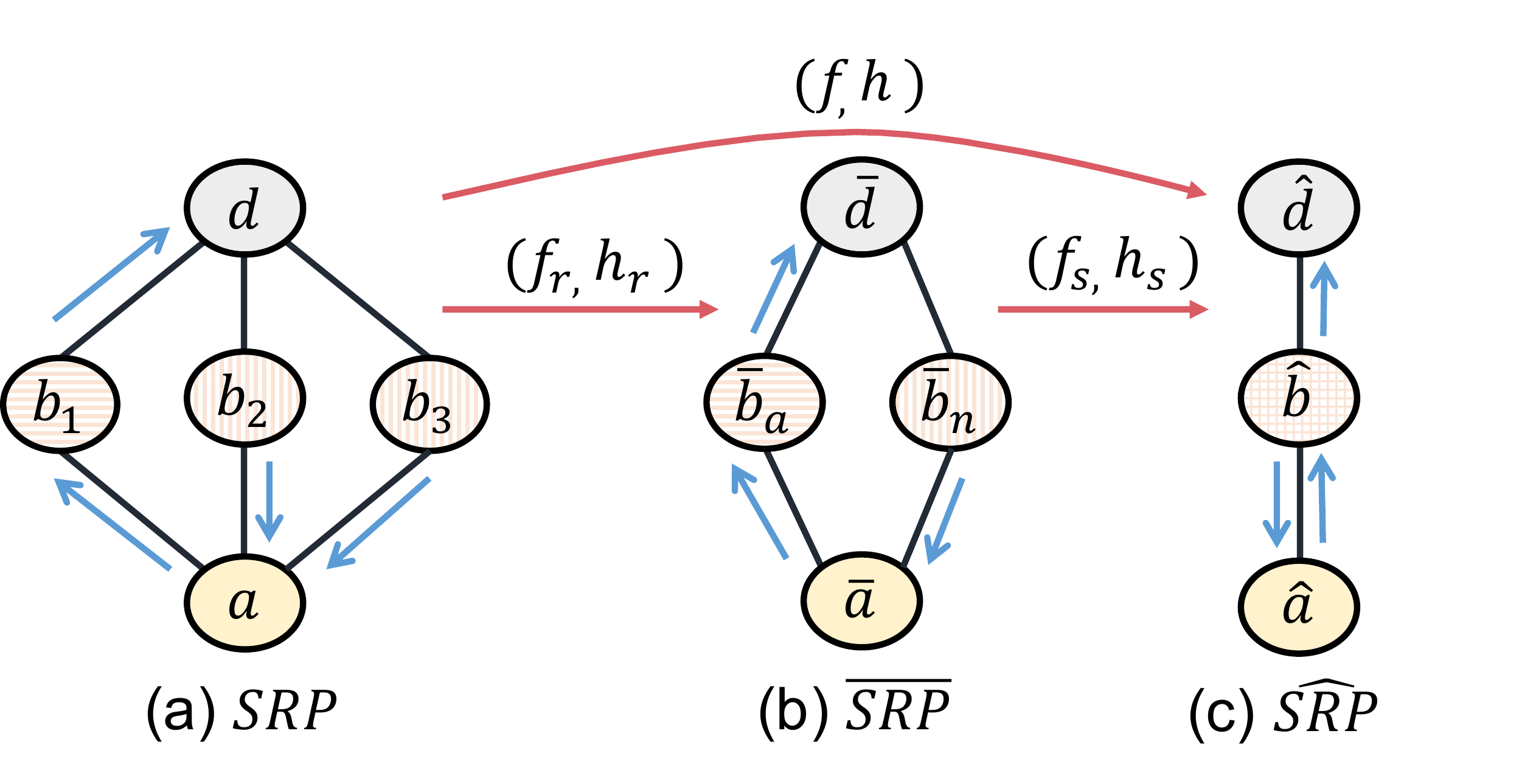}
  \vspace{-1em}
  \caption{Abstraction refinement for  Figure~\ref{fig:bgp-abstraction}(a).}
  \label{fig:refinement}
\end{figure}

\para{Abstraction refinement}
\label{ssec:refinement}
A bound on the the number of behaviors for nodes
in BGP lets us refine an abstraction by splitting apart
abstract nodes into enough cases to recover
\cpequivalence. 
%
We now formalize this intuition.

Suppose we are given an $SRP = (G,A,a_{d},\prec,\transfer)$ for BGP and its abstract version $\wh{SRP} = (\wh{G}, \wh{A},\wh{a_{d}}, \wprec, \wh{\transfer})$, which are self-loop-free
and created from a \aaabs $(f,h)$. We define a new abstraction $\ol{SRP} = (\ol{G}, \ol{A}, \ol{a_{d}}, \oprec, \ol{\transfer})$ obtained by splitting up each node $\wh{v}$ into $\abs{\prefs(\wh{v})}$ copies of the node.
We can view the mapping from $SRP$ to $\wh{SRP}$ as the composition of two abstractions $(f_r, h_r)$ from $SRP$ to $\ol{SRP}$, and $(f_s, h_s)$ from $\ol{SRP}$ to $\wh{SRP}$, where the comparison and transfer functions for $\ol{SRP}$ are copied from $\wh{SRP}$.
Given a new abstraction $(f_r, h_r)$ where $f_r : V \rightarrow \ol{V}$ and $h_r : A \rightarrow \ol{A}$, we say $(f_r,h_r)$ \emph{refines} $(f,h)$, written as $(f_r,h_r) \sqsubseteq_{(f_s,h_s)} (f,h)$ if $f_r$ is an \emph{onto} function, and $f = f_s \circ f_r$ and $h = h_s \circ h_r$.

We now show that there is a bisimulation between the solutions $\lab$ and $\olab$ as before. However, whereas the abstraction mapping $f$ was known in advance, the refined mapping $f_r$ may change depending on the particular solution $\lab$. 
For example, Figure~\ref{fig:refinement}(a) shows one of three possible forwarding behaviors for the network. As discussed earlier, with a different message arrival timing, other solutions would have emerged.
Depending upon this solution, different nodes, \EG $\{b_1, b_2\}$
or $\{b_1, b_3\}$ would be mapped to $\ol{b_n}$.  We do not know which concrete
nodes are mapped to which abstract nodes, but we do know that the abstraction has sufficiently many nodes to characterize all possible behaviors.

{\iffull
\begin{theorem} \label{thm:refinement-equivalence}
  Suppose we have well-formed $SRP$, $\wh{SRP}$, and $\ol{SRP}$ for BGP with an \effective abstraction $(f, h)$. For any solution $\lab$ to $SRP$, there exists a refinement $(f_r, h_r) \sqsubseteq_{(f_s,h_s)} (f, h)$ where $\ol{\lab}$ is a solution to $\ol{SRP}$, and $\lab$ and $\ol{\lab}$ are label- and fwd-equivalent.

  {\iffull
  \begin{proof}
    First, we will show a particular refinement.
    From Theorem~\ref{thm:bounded-behaviors}, we know that any solution to $\lab$ can only have $|\prefs(\wh{v})| + 1$ behaviors.
    Let use define $f_r(v) = \ol{v} = \behavior(f(v))(i)$, where this notation means that we pick out the $i$th node in $\ol{V}$ such that $f_s$ maps it to $\wh{v}$.
    We can modify this scheme slightly to ensure that $f_r$ is an \emph{onto} function, if no node would map to the $k$th behavior, then we pick an arbitrary node that maps to the $j$th behavior (if there is more than one node that maps to the $j$th behavior), and map it to the $k$th behavior instead.
    This is a valid refinement to $(f,h)$ since $f = f_s \circ f_r$ and $f_r$ is onto.
    
    Since we have a particular solution $\lab$ that is loop-free (since BGP is loop-free), we know all the edges in $SRP$ that are not used due to loops. 
    For example, in Figure~\ref{fig:refinement} in the concrete network (left), the green node would have transfer function $\bot$ from each of the red neighbors below due to loop prevention.
    
    Consider an isomorphic network $G'$, where all such edges are removed (e.g., directed edges from the green to red nodes). Similarly, in the refined network $\ol{G}'$, we would remove the corresponding edges (e.g., the directed edge from the green to red nodes).

    The particular refinement $f_r$ we chose is important because we will still have an \aaabs after removing these edges since each node with such unique behavior rejected the same nodes (due to loops) to accept the worse path. Therefore, removing the abstract edge and concrete edges remains a universal abstraction.

    The same solution $\lab$ is a solution for the isomorphic $SRP$ where the loop-prevention mechanism for BGP is removed (i.e., we don't block paths with loops). Since we have relative-transfer-equivalence, by removing the loop condition, we get full transfer-equivalence. We can then simply appeal to Theorem~\ref{thm:correctness-local} to derive \eafwd and preference-equivalence of $\lab$ and $\ol{\lab}$ for the isomorphic networks.

    Finally, if we add back the abstract edges that we removed, we need to show that we still have the same solution $\ol{\lab}$ with loop-prevention. We do this by showing that such edges would be rejected as loops.
    Given that we have \cpequivalence, and in the concrete solution $\lab$ this edge would result in a loop of the form $u,w_1,\ldots,w_k,u$, we know that the abstract path would also have a loop $f_r(u),f_r(w_1),\ldots,f_r(w_k),f_r(u).$ 
  \end{proof}
  \fi}
\end{theorem}

Next we show the other direction. Note that in both cases, the proof is constructive and thus responsible for identifying the particular appropriate refinement $(f_r, h_r)$ and $(f_s, h_s)$.

\begin{theorem} \label{thm:refinement-equivalence2}
  Suppose we have well-formed $SRP$, $\wh{SRP}$, and $\ol{SRP}$ for BGP with an \effective abstraction $(f, h)$. For any solution $\ol{\lab}$ to $\ol{SRP}$, then there exists a refinement $(f_r, h_r) \sqsubseteq_{(f_s,h_s)} (f, h)$ where $\lab$ is a solution to $SRP$, and $\lab$ and $\ol{\lab}$ are label- and fwd-equivalent.

  \begin{proof}
    Setup $f_r$ such that, for each node $\ol{v}$ that forwards for an attribute that is not the best when ignoring loop-prevention, we have a single node in $v \in V$ map to such a $\ol{v}$. For every other node $\ol{v}$, that forwards to the best available option, we map every other $v$ to each of these $\ol{v}$. That is, we assign a single concrete node for each unique behavior that is not the best route and all other nodes map to the abstract node that has the best route.

    As before, we remove each edge in $\ol{SRP}$ that corresponds to an edge rejected due to loops in $\ol{\lab}$, and all corresponding concrete edges related under $f_r$. As before, in the concrete network, we will still have a \aaabs since each concrete node that forwards to a non-best path does so because the better paths are rejected due to loops.

    This network will have the same solution $\ol{\lab}$ but has full transfer-equivalence, and we can again appeal to Theorem~\ref{thm:correctness-local} for preference- and \cpequivalence for BGP without loop-prevention.

    If we add back the abstract edges that we removed, we need to show that the same solution $\lab$ is still a solution with loop-prevention. Suppose that the abstract node $\ol{u}$ prevented a loop $\ol{u},\ol{w}_1,\ldots,\ol{w}_k,\ol{u}$.
    Then each node $u$ where $f_r(u) = \ol{u}$ that chose a non-best path in the concrete network also did so due to loop-prevention.
  \end{proof}

\end{theorem}

\else 

\begin{theorem} \label{thm:refinement-equivalence}
  Suppose we have well-formed $SRP$, $\wh{SRP}$, and $\ol{SRP}$ for BGP with an \effective abstraction $(f, h)$. There is a solution $\lab$ to $SRP$ iff there is a solution $\ol{\lab}$ to $\ol{SRP}$, such that there exists a refinement $(f_r, h_r) \sqsubseteq_{(f_s,h_s)} (f, h)$ and $\lab$ and $\ol{\lab}$ are label- and fwd-equivalent.
\end{theorem}
\fi}

\OMIT{
Finally, we have the refinement analogue
to Corollary~\ref{thm:all-protocol-equiv}.

\begin{corollary} \label{thm:bgp-equiv}
  Suppose we have well-formed $SRP$, $\wh{SRP}$, and $\ol{SRP}$ for BGP with an \effective abstraction $(f,h)$. There is a solution $\lab$, where each node $u_1 \mapsto \wh{u_1}$ forwards along path $s = \lab(u_1) \ldots \lab(u_k)$ to some node $u_k \mapsto \wh{u_k}$ \ifft there is a solution $\ol{\lab}$ where each node $\ol{u_1} \mapsto \wh{u_1}$ forwards along path $\ol{s} = \lab(\ol{u_1}) \ldots \lab(\ol{u_k})$ to some $\ol{u_k} \mapsto \wh{u_k}$ such that $h(s) = h_s(\ol{s})$.

  {\iffull
  \begin{proof} We show each direction separately.

    \vspace{.8em}
    \noindent \textbf{Case ($\Rightarrow$)}
    Suppose $\lab$ is a solution for $SRP$. From Theorem~\ref{thm:refinement-equivalence}, we know there exists a refinement $(f_r, h_r) \sqsubseteq (f,h)$ of for $\ol{SRP}$ with solution $\ol{\lab}$, and also that $\lab$ and $\ol{\lab}$ are fwd-equivalent. 
    Given any two nodes $u$ and $v$ where $u$ can reach $v$, there exists a path $p = u,w_1,\ldots,w_k,v$ where $(u,w_1) \in \fwd(u)$ and $(w_i,w_{i+1}) \in \fwd(w_i)$ and $(w_k,v) \in \fwd(w_k)$. Because $\lab$ and $\ol{\lab}$ are fwd-equivalent, we know that $(f_r(u),f_r(w_1)) \in \ol{\fwd}(f_r(u))$ and so on. 
    Therefore, there is an abstract path in $\ol{\lab}$ where $f_r(u)$ can reach $f_r(v)$ of the form $f_r(u), f_r(w_1), \ldots, f_r(w_k), f_r(v)$. Since $f_r$ is \emph{onto} from Theorem~\ref{thm:refinement-equivalence}, we know that this is the case for every $f_r(u) \in f_s^{-1}(f(u))$. 
    Observe that the labels of the concrete path are $s = \lab(u), \lab(w_1), \ldots, \lab(w_k), \lab(v)$. Similarly, the abstract path has labels $\ol{\lab}(f_r(u)), \ldots, \ol{\lab}(f_r(v))$. 
    Due to label-equivalence, this is the same as $h_r(\lab(u)), \ldots, h_r(\lab(v))$, which is just $h_r(s)$.
    
    Recall that we must show that $h_s(\ol{s}) = h(s)$
    Since we know $h_r(s) = \ol{s}$, we have $h_s(h_r(s)) = h(s)$. Finally, because $h = h_s \circ h_r$, these are equivalent.

    \vspace{.8em}
    \noindent \textbf{Case ($\Leftarrow$)}
    Suppose $\ol{L}$ is a solution for $\ol{SRP}$, then from Theorem~\ref{thm:refinement-equivalence}, we know that  there exists an onto refinement $(f_r,h_r) \sqsubseteq (f,h)$ where $\lab$ and $\ol{\lab}$ are \eafwd. Consider an arbitrary path $\ol{u},\ol{w_1},\ldots,\ol{w_k},\ol{v}$. Then we know $(\ol{u},\ol{w_1}) \in \ol{\fwd}(\ol{u})$ and so on. 
    From the fact that $\lab$ and $\ol{\lab}$ are \eafwd, every node $u$ that maps to $\ol{u}$ forwards to the same neighbor. That is, we know that each $u$ where $f_r(u) = \ol{u}$, has the same $w_1$ where $f_r(w_1) = \ol{w_1}$ and $(u,w_1) \in \fwd(u)$ and so on. 
    Therefore, each node $u \in f_r^{-1}(\ol{u})$ has the same path starting after $w_1$: $u,w_1,\ldots,w_k,v$. The abstract path has labels $\ol{\lab}(f_r(u)),\ldots,\ol{\lab}(f_r(v))$. Due to label-equivalence, this is the same as $h_r(\lab(u)),\ldots,h_r(\lab(v))$, which is $h_r(s)$. 
    
    Once again, we have $h_s(h_r(s)) = h(s)$, which follows from the fact that $h = h_s \circ h_r$.
  \end{proof}
  \fi}

\end{corollary}
}

A key difference between Theorem~\ref{thm:correctness-local} and
Theorem~\ref{thm:refinement-equivalence}, is that the forwarding paths between the
concrete network ($SRP$) and the refined network ($\ol{SRP}$) will only be equivalent with respect to the
original abstract network ($\wh{SRP}$). For example, in
Figure~\ref{fig:refinement}(a), if we want to check that $b_2$ and
$b_3$ forward along a path that satisfies some property
$p$, then we can not check it against only $\ol{b_a}$ in Figure~\ref{fig:refinement}(b).
Rather, we have to check it against $\ol{b_n}$ as well because there is another stable solution where
the roles of $b_a$ and $b_n$ are reversed.

\subsection{Properties preserved}
\label{ssec:consequences}
\hl{As a consequence of} \CPequivalence, \hl{for a solution} $\hlmath{\lab}$ \hl{to the concrete network, there exists a path} $\hlmath{s = x_1, \ldots, x_k}$ \hl{where the network forwards along} $\hlmath{s}$ \hl{with labels} $\hlmath{\lab(x_1), \ldots, \lab(x_k)}$ \hl{iff for some solution} $\hlmath{\wlab}$ \hl{to the abstract network, there is a path} $\hlmath{f(s)}$ \hl{where the abstract network forwards along} $\hlmath{f(s)}$ \hl{with labels} $\hlmath{\wlab(f(x_1)), \ldots, \wlab(f(x_k))}$.
%
%
%
%
\hl{Concretely, one can check that any of the following properties hold on small abstract networks and be sure the concrete counterpart satisfies the property as well.}

\begin{itemize}[leftmargin=*]
\item{\bf Reachability:} $\hlmath{f(u)}$ \hl{can reach} $\hlmath{f(v)}$ \hl{in the abstract network iff} $\hlmath{u}$ \hl{can reach} $\hlmath{v}$ \hl{in the concrete network.}

\item{\bf Path Length:}  \hl{All paths between} $\hlmath{f(u)}$ \hl{and} $\hlmath{f(v)}$ \hl{have length} $\hlmath{n}$ \hl{iff all paths between} $\hlmath{u}$ \hl{and} $\hlmath{v}$ \hl{have length} $\hlmath{n}$.

\item{\bf Black Holes:}  \hl{Path} $s$ \hl{in the concrete network ends with label} $\hlmath{\bot}$ \hl{iff path} $f(s)$ \hl{ends with} $\hlmath{\bot}$ \hl{in the abstract network}.

\item{\bf Multipath Consistency:}
  \hl{Traffic sent from} $\hlmath{f(u)}$ \hl{is reachable along some path to} $\hlmath{f(v)}$ \hl{but dropped along another path iff traffic from} $\hlmath{u}$ \hl{is
  reachable along some path to} $\hlmath{v}$ \hl{and dropped along another path.}

\item{\bf Waypointing:}
  \hl{Traffic will be waypointed through one of}
  $\hlmath{\{ f(w_1), \ldots, f(w_n) \}}$ \hl{in the abstract network iff it will go through one of} $\hlmath{\{ w_1, \ldots, w_n \}}$ \hl{in the concrete network.}

\item{\bf Routing Loops:}
  \hl{There is a routing loop in the abstract network iff there is one in the concrete network.}
%
\end{itemize}

\para{Convergence}
\hl{The concrete network necessarily diverges (has no stable solution) iff the abstract network necessarily diverges. To see why, suppose the concrete network had no stable solution, but the abstract network had a stable solution. This would violate} \CPequivalence, \hl{since each abstract solution has a corresponding concrete solution. Similarly, the concrete network can converge (has some stable solution) iff the abstract network can converge. However,} \CPequivalence \hl{alone does not guarantee that networks that might converge or might diverge, like the naughty gadget in BGP}~\cite{stable-paths}, \hl{will necessarily reduce to an abstract network that may diverge.} 

\hl{On the other hand,} \effective \hl{abstractions are stronger than (imply)} \CPequivalence.  \hl{We postulate (but have not proven) that an} \effective \hl{abstraction is sufficient to preserve convergence. For example, it would appear that the concrete network will have a dispute wheel}~\cite{stable-paths} \hl{(the lack of which is sufficient condition for convergence safety and robustness) iff the abstract network has a dispute wheel (the nodes in concrete network forming a dispute wheel will induce a dispute wheel in their abstract counterpart).}

\subsection{Properties not preserved}

While effective abstractions preserve the nature of forwarding paths, they do not, in general, preserve the number of paths or the number of neighbors.  Indeed, that is the point---effective abstractions usually reduce the number of paths  and neighbors to speed analysis. \hl{Consequently, we cannot reason faithfully about properties such as fault tolerance, load balancing, or any QoS properties. For instance, in the abstract network, a single link failure may partition a network whereas in the concrete network, there may exist two or more link-disjoint paths between all pairs of nodes, allowing the concrete network to tolerate any single failure.} 
%

\OMIT{
\begin{figure}[t]\small

  \hrulefill  
  
  \begin{minipage}[t]{1\linewidth}
  \hdr{Regex Syntax}{\quad \fbox{$R(A)$}}
  %
  \[ \begin{array}{lclr}
    p &:& A \rightarrow \{0,1\} & \textit{First-order Predicate} \\
    r &::=& p ~\vert~ r_1 + r_2 ~\vert~ r_1 \cdot r_2 ~\vert~ r^*  & \textit{Regular expression} \\
  \end{array} \] 
  
  \vspace{1em}
  \hdr{Regex Matching}{\quad \fbox{$r \sim s$}}  
  \[
    \begin{array}{lcl}
      p \sim a              &\iff& p(a) \\
      r_1 + r_2 \sim s      &\iff& r_1 \sim s ~\vee~ r_2 \sim s   \\
      r_1 \cdot r_2 \sim s  &\iff& \exists i.~ r_1 \sim s_0..s_i ~\wedge~ r_2 \sim s_{i+1}..s_n \\
      r^* \sim \emptyset \\
      r^* \sim s            &\iff& r \cdot r^* \sim s \\
    \end{array}
    \] 

    \vspace{1em}
    \hdr{Regex Abstraction}{\quad \fbox{$f(r)$}}  
    \[
      \begin{array}{lcl}
        f(p)                  &=& p[v_i / f(v_i)] \\
        f(r_1 + r_2)          &=& f(r_1) + f(r_2) \\
      \end{array}
      \hspace{2em}
      \begin{array}{lcl}
        f(r_1 \cdot r_2)      &=& f(r_1) \cdot f(r_2) \\
        f(r^*)                &=& f(r)^* \\
      \end{array}
      \] 
  \end{minipage}

  \hrulefill 

  \caption{Property language syntax and semantics.}
  \label{fig:property-language}
\end{figure}
}

%
%
%
%

\OMIT{
\section{Property Language}

Corollaries \ref{thm:all-protocol-equiv} and \ref{thm:bgp-equiv}
establish tha the forwarding paths
of the concrete and abstract networks coincide, but they
do not define an interface network operators can use to
investigate properties of their networks.  In this section, we introduce a
operator query language based on regular expressions over paths of
attributes.  With this language in hand, operators can ask a wide
variety of questions about their networks, including questions about
data plane forwarding, such as reachability, as well as questions about
control plane dynamics, such as whether announcements leaving the operator's
network are tagged with the right community values.

Our language, which is parameterized over a particular set of attributes $A$, is shown in Figure~\ref{fig:property-language}. We let a predicate be any first-order property over attributes $A$ and let $R(A)$ denote regular expressions over these predicates. While it may seem limiting to only allow expressions over attributes rather than topology nodes, it is easy to add nodes to the attribute. Given an SRP with nodes $V$, we can always construct a new problem with attributes $A' = A \times \mathrm{list}(V)$, a ranking function that only ranks the original attribute, and a transfer function that appends the current node to list of vertices (similar to BGP's AS path). This list is used only to help ask questions about the network and does not affect the underlying protocol.

The semantics of regular expressions is defined over a sequence of attributes $s = a_1, \ldots, a_n$. We use $\emptyset$ to denote the empty sequence and $s_i..s_j$ to denote the subsequence of $s$ from element $i$ to $j$.
The (standard) matching relation ($r \sim s$) defines the paths $s$ that match
regular expression $r$.
Finally, to abstract a regular expression, we replace all instances of predicate $p$ with a new predicate $p[v_i / f(v_i)]$, which replaces all references to $v_i$ with $f(v_i)$.

\subsection{Property Preservation}
\label{sec:language}

Given a property in our language, we want to guarantee that the property holds for the concrete network \ifft it holds for the abstract network. 

\begin{defn}
  We say that the abstraction is predicate-equivalent with respect to abstraction $(f,h)$ for a predicate $p$ if 
  $$\forall a.~ p(a) \iff f(p)(h(a))$$
\end{defn}  

{\iffull
\begin{theorem} \label{thm:property-abstraction}
  If regex $r$ is predicate-equivalent with respect to an abstraction $(f,h)$, then for any path $r \sim s \iff f(r) \sim h(s)$.

  \begin{proof} By induction on the lexicographical order of $r$ followed by the length of $s$.

  \noindent \textbf{(Case $p$)} 
  By applying the definition of $\sim$, and predicate-equivalence, we get:
  $p \sim a \iff p(a) \iff p[v_i/f(v_i)] \sim h(a) \iff f(p) \sim h(a)$.

  \noindent \textbf{(Case $r_1 + r_2$)} We know 
  $r_1 + r_2 \sim s \iff r_1 \sim s$ or $r_2 \sim s$. if $r_1 \sim s$, then by induction, $f(r_1) \sim h(s)$ and therefore $f(r_1 + r_2) \sim h(s)$. The case for $r_2$ is symmetric.

  \noindent \textbf{(Case $r_1 \cdot r_2$)}
  $r_1 \cdot r_2 \sim s \iff \exists i.~ r_1 \sim s_0..s_i \wedge r_2 \sim s_{i+1}..s_n$. By induction, suppose such an $i$ exists. The IH tells us that $f(r_1) \sim h(s_0..s_i) = h(s_0)..h(s_i)$ and $f(r_2) \sim h(s_{i+1}..s_n) = h(s_{i+1})..h(s_n)$.
  We then know that:
  $f(r_1 \cdot r_2) = f(r_1) \cdot f(r_2) \sim h(s)$

  \noindent \textbf{(Case $r^*$)}
  We know $r^* \sim s$. There are two cases. If $s = \emptyset$, then we know $r^* \sim \emptyset$ and $h(s) = \emptyset$ and thus $f(r^*) = f(r)^* \sim \emptyset$. In the case where $s$ is non-empty, we know that $r^* \sim s \iff r \cdot r^* \sim s$.
  By induction ($s$ getting smaller), we know that this happens \ifft $f(r) \cdot f(r)^* \sim h(s)$. Since we also know that $f(r) \cdot f(r)^* \sim h(s) \iff f(r)^* \sim h(s) \iff f(r^*) \sim s$, the proof is complete.
  \end{proof}
\end{theorem}
\fi}

\begin{theorem} \label{thm:bgp-regex-equiv}
  Suppose we are given self-loop-free $SRP$, $\ol{SRP}$, and $\wh{SRP}$ along with regular property $r$ that is predicate-equivalent with respect to \effective abstraction $(f,h)$.
  There is a solution $\lab$, where each node $u \mapsto \wh{u}$ has a path $(s = \lab(u) \ldots \lab(v))$ to some node $v \mapsto \wh{v}$ such that $r \sim s$ \ifft there is a solution $\ol{\lab}$ where each node $\ol{u} \mapsto \wh{u}$ has a path $\ol{s}$ to some $\ol{v} \mapsto \wh{v}$ such that $f(r) \sim h_s(\ol{s})$.

  {\iffull
  \begin{proof} We show each direction separately.

    \vspace{.8em}
    \noindent \textbf{Case ($\Rightarrow$)}
    Suppose $\lab$ is a solution for $SRP$. From Theorem~\ref{thm:refinement-equivalence}, we know there exists a refinement $(f_r, h_r) \sqsubseteq (f,h)$ of for $\ol{SRP}$ with solution $\ol{\lab}$, and also that $\lab$ and $\ol{\lab}$ are \eafwd. 
    Given any two nodes $u$ and $v$ where $u$ can reach $v$, there exists a path $p = u,w_1,\ldots,w_k,v$ where $(u,w_1) \in \fwd(u)$ and $(w_i,w_{i+1}) \in \fwd(w_i)$ and $(w_k,v) \in \fwd(w_k)$. Because $\lab$ and $\ol{\lab}$ are \eafwd, we know that $(f_r(u),f_r(w_1)) \in \ol{\fwd}(f_r(u))$ and so on. 
    Therefore, there is an abstract path in $\ol{\lab}$ where $f_r(u)$ can reach $f_r(v)$ of the form $f_r(u), f_r(w_1), \ldots, f_r(w_k), f_r(v)$. Since $f_r$ is \emph{onto} from Theorem~\ref{thm:refinement-equivalence}, we know that this is the case for every $f_r(u) \in f_s^{-1}(f(u))$. 
    Observe that the labels of the concrete path are $s = \lab(u), \lab(w_1), \ldots, \lab(w_k), \lab(v)$. Similarly, the abstract path has labels $\ol{\lab}(f_r(u)), \ldots, \ol{\lab}(f_r(v))$. 
    Due to preference-equivalence, this is the same as $h_r(\lab)(u)), \ldots, h_r(\lab)(v))$, which is just $h_r(s)$.

    Now, we need to show that $f(r) \sim h_s(\ol{s})$ assuming that $r \sim s$. Since we know $h_r(s) = \ol{s}$, we must show that $f(r) \sim h_s(h_r(s))$. However, since $h = h_s \circ h_r$, this is the same as showing $f(r) \sim h(s)$. This follows from Theorem~\ref{thm:property-abstraction}.

    \vspace{.8em}
    \noindent \textbf{Case ($\Leftarrow$)}
    Suppose $\ol{L}$ is a solution for $\ol{SRP}$, then from Theorem~\ref{thm:refinement-equivalence}, we know that  there exists an onto refinement $(f_r,h_r) \sqsubseteq (f,h)$ where $\lab$ and $\ol{\lab}$ are \eafwd. Consider an arbitrary path $\ol{u},\ol{w_1},\ldots,\ol{w_k},\ol{v}$. Then we know $(\ol{u},\ol{w_1}) \in \ol{\fwd}(\ol{u})$ and so on. 
    From the fact that $\lab$ and $\ol{\lab}$ are \eafwd, every node $u$ that maps to $\ol{u}$ forwards to the same neighbor. That is, we know that each $u$ where $f_r(u) = \ol{u}$, has the same $w_1$ where $f_r(w_1) = \ol{w_1}$ and $(u,w_1) \in \fwd(u)$ and so on. 
    Therefore, each node $u \in f_r^{-1}(\ol{u})$ has the same path starting after $w_1$: $u,w_1,\ldots,w_k,v$. The abstract path has labels $\ol{\lab}(f_r(u)),\ldots,\ol{\lab}(f_r(v))$. Due to preference-equivalence, this is the same as $h_r(\lab(u)),\ldots,h_r(\lab(v))$, which is $h_r(s)$. We know that $f(r) \sim h_s(\ol(s)) = h_s(h_r(s)) = h(s)$. From Theorem~\ref{thm:property-abstraction}, this means that $r \sim s$.
  \end{proof}
  \fi}

\end{theorem}

This theorem is a direct consequence of Theorems~\ref{thm:all-protocol-equiv} and~\ref{thm:bgp-equiv}. We state it using the more general case with a refinement $\ol{SRP}$, but settings without BGP can be trivially captured with the identity refinement (\IE, $\ol{SRP}$ and $\wh{SRP}$ are the same).
 
\subsection{Example Properties}
\label{sec:example-properties}
We now look at some example properties that satisfy the predicate-equivalence constraint. \dpw{Given a regular expression $r$, what is the network verification
  problem?  Likely: for all $\lab$, for all $s \in \lab$, $r \sim s$.  Have
  we implemented that?  Explain.  Also:  I added a no export example to
  illustrate reasoning about the control plane for ratul.}

\para{Reachability}
We can model reachability to the destination $d$ (\EG, to answer questions like ``are two hosts reachable'') using a single predicate $\ptrue$, which returns true for any attribute. This trivially satisfies predicate-equivalence since $f(\ptrue) = \ptrue$ and returns 1 on any attribute $a$ (or $h(a)$).
We can encode reachability between $\wh{u}$ and $\wh{d}$ as the existence of a path that matches the regular expression ${\ptrue}^*$. That is, if there is a path $\wh{s}$ such that ${\ptrue}^* \sim \wh{s}$ from $\wh{u}$ to $\wh{d}$, each concrete node mapping to $\wh{u}$ has a path $s$ to $d$ where ${\ptrue}^* \sim s$.

\para{Path Length}
We can check path length similarly. For example, to check if a concrete node has a path of length 3, we can use the regular expression $\ptrue \cdot \ptrue \cdot \ptrue$. Similarly, if there is no abstract path that has a length of 4 or higher, then there is not a concrete path that has a length of 4 or higher, and so on, since we can relate them to regular expressions.

\para{Black holes}
We can check for black holes by introducing a predicate that matches the $\bot$ attribute.
Let $\pbot$ be a predicate that returns true if the attribute is $\bot$ and false otherwise. Since we know that $h(\bot) = \bot$ for any abstraction, we get predicate-abstraction for free.
We can encode black holes at some node as a path to that node that ends with the $\bot$ label, \IE, we use the expression $({\ptrue}^* \cdot \pbot)$

\para{Mulitpath Consistency}
We can check if traffic that is forwarded along multiple paths is dropped along some but not others.
If there is a path between $\wh{u}$ and $\wh{v}$, and a second path from $\wh{u}$ that is dropped at some other node $\wh{v'}$ before getting to $\wh{v}$, then we can capture this with two expressions, one for reachability to $\wh{v}$ and a second for a black hole to $\wh{v'}$. This holds in the abstract network \ifft two such paths exist in the concrete network.

\para{Waypointing}
Without loss of generality, we can include the actual router in the attribute $A$ and have the transfer function set the new router in the attribute.
Let the predicate $p_X$ denote a hop at router $X$. That is, the function $p_X(a) = a.{\mathrm{router}} = X$. Suppose now we want to ask if traffic will be waypointed through one of a collection of routers $\{ w_1, \ldots, w_n \}$. Such a property can be encoded using the regular expression: ${p_{\mathrm{true}}}^* \cdot (p_{w_1} + \dots + p_{w_n}) \cdot {p_{\mathrm{true}}}^*$. From Theorem~\ref{thm:bgp-regex-equiv}, we know that this is the case iff there is an abstract path that matches regex ${p_{\mathrm{true}}}^* \cdot (p_{f(w_1)} + \dots + p_{f(w_n)}) \cdot {p_{\mathrm{true}}}^*$.


\vspace{.8em}
\noindent \textbf{Routing Loops.}
Routing loops are an interesting case. We can describe a loop from some node $\wh{u}$ to itself using the regular expression $p_{\wh{u}} \cdot .^+ \cdot p_{\wh{u}}$. However this only guarantees us that traffic sent from $u \mapsto \wh{u}$ will eventually reach some node $u' \mapsto \wh{u}$. However $u$ and $u'$ may not necessarily be the same node. 
However, because there are a finite number of concrete nodes $u' \mapsto \wh{u}$, and because each such $u'$ must eventually reach back to another node that maps to $\wh{u}$, eventually there will be a concrete routing loop.

\vspace{.8em}
\noindent \textbf{No Export.}
There are a wide variety of properties one might wish to verify about
the control plane passes announcements around.  One simple example is
a ``no export'' policy---announcements tagged with
the $\mathsf{noExport}$ community do not leave the operator's network.
Now, assuming we have
predicates $\mathsf{in}$ (we own the given router) and $\mathsf{noExport}$
(the current attribute is tagged with the $\mathsf{noExport}$
community), the regular expression
$\mathsf{noExport} . \mathsf{in}^*$ identifies paths that are tagged
 $\mathsf{noExport}$ and remain within the network.
}

%
%
%
%

\section{Abstraction Algorithm}
\label{sec:algorithm}

Earlier sections described the conditions under which an abstraction will preserve \cpequivalence, but they give no insight into how one might compute such an abstraction. In this section, we describe an algorithm that computes an abstraction directly from a set of router configurations.

\subsection{Algorithm Overview}

Our algorithm starts with the following observations.
The key requirement for computing an \effective abstraction is to ensure that we satisfy each required condition in Figure~\ref{fig:spp-and-abstraction}.  
Some conditions such as orig-equivalence ($h(a_d) = \wh{a_d}$), drop-equivalence ($h(a) = \bot \iff a = \bot$) and rank-equivalence ($a \prec b \iff h(a) ~\wprec~ h(b)$) depend only on the particular protocol and choice of $h$. By fixing $h$ in advance for each protocol similar to those used in Figures~\ref{fig:srp-bgp} and \ref{fig:srp-static}, we can guarantee that these conditions hold regardless of the configurations.
Other conditions such as dest-equivalence and \aeabs depend on the topology, but not the policy embedded in configurations.

Transfer-equivalence: $h(\transfer(e,a)) = \wtransfer(f(e), h(a))$ is the only condition that depends on user-defined policy. Suppose two concrete edges $e_1$ and $e_2$ are mapped together by the topology function $f$. We would have $h(\transfer(e_1,a)) = \wtransfer(f(e_1), h(a)) = \wtransfer(f(e_2), h(a)) = h(\transfer(e_2,a))$. One simple way to ensure that this equality holds is to only combine together nodes with the same transfer function. In our example, $\transfer(e_1,a) = \transfer(e_2,a)$ would suffice to allow $e_1$ and $e_2$ to map to the same abstract edge.

Based on the observations above, we fix $h$; our remaining task is to find a suitable $f$ that satisfies the topology abstraction requirements and only maps together edges with equivalent transfer functions (for the destination $d$). We find such a function $f$ using an algorithm based on abstraction refinement. We start with the coarsest possible abstraction where there is a single abstract destination node $\wh{d}$ and one other abstract node for all other concrete nodes, and while the abstraction violates $\aeabs$ or transfer-equivalence, we refine it by breaking up the problematic abstract node into multiple abstract nodes. 


   
\begin{figure}
  \includegraphics[scale=.28]{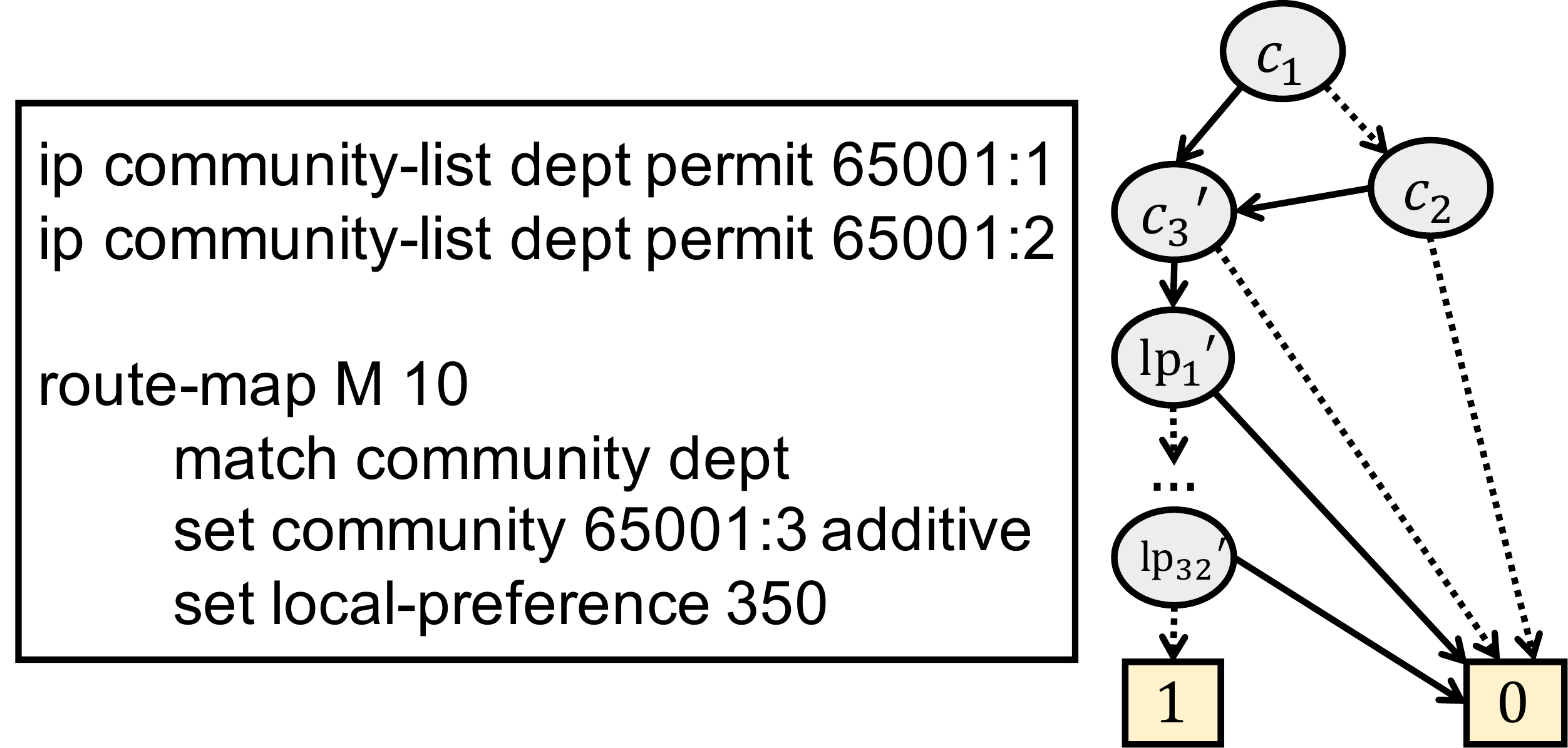}
  \vspace{-.7em}
  \caption{BDD for a BGP policy on an interface.}
  \label{fig:bdd-example} 
\end{figure} 

For efficiency, before abstraction refinement, we process router configurations in two different ways.

\para{1. Destination Equivalence Classes (ECs)}
In our theoretical account of routing, each SRP contains a single destination.
However, in practice, configurations contain routing information for many destinations
simultaneously.  Because announcements for (most) destinations do not
interact with one another,
we can partition the network into equivalence classes based on where destinations are rooted.
Each class has a collection of destination IP ranges and destination node(s). This partitioning allows us to build one abstraction per class instead of one per address. To partition the network into equivalence classes, we use a prefix-trie data structure where the leaves of the trie contain a set of destination nodes.

\begin{algorithm}[t]
  \begin{footnotesize}
  \caption{Compute abstraction function $f$} \label{alg:alg1}
  \begin{algorithmic}[1]
  
  \Procedure{FindAbstraction}{Graph G, Bdds bdds}
  \State \Call{Specialize}{bdds, G.d}
  \State f $\gets$ \Call{UnionSplitFind}{G.V}
  \State \Call{Split}{f, $\{G.d\}$}
  \While{True}
    \State $\wh{V}$ $\gets$ \Call{Partitions}{f}
    \For{ $\wh{u}$ \textbf{in} $\wh{V}$ }
      \If{$\vert \wh{u} \vert$ $\leq$ 1} \textbf{continue}
      \EndIf
      \State \Call{Refine}{G, bdds, f, $\wh{u}$, $\vert$prefs($\wh{u}$)$\vert$}  
    \EndFor
  
    \State $\wh{V}' \gets$ \Call{Partitions}{f}  
    \If{$\vert \wh{V} \vert$ = $\vert \wh{V}' \vert$} \textbf{break}
    \EndIf
  
  \EndWhile
  \State \textbf{return} \Call{SplitIntoBGPCases}{f}
  \EndProcedure
  
  \State
  
  \Procedure{Refine}{G, bdds, f, $\wh{u}$, numPrefs}
  \State map $\gets$ \Call{CreateMap}{}
  \For{$u$ $\in$ $\wh{u}$}
    \For{$e = (u,v)$ $\in$ G.E}
      \State pol $\gets$ \Call{Get}{bdds, e}
      \State n $\gets$ (numPrefs > 1 $?$ $v$ $:$ $f(v)$)
      \State map[u] $\gets$ map[u] $\cup$ $\{$ (pol, n) $\}$
    \EndFor
  \EndFor
  \For{us $\in$ \Call{GroupKeysByValue}{map}} 
    \State \Call{Split}{f,us}
  \EndFor
  \EndProcedure
  \end{algorithmic}
  \end{footnotesize}
\end{algorithm}

\para{2. Encoding transfer function using BDDs}
In order to efficiently find all interfaces that have equivalent transfer-functions for a given destination (class), we use Binary Decision Diagrams (BDDs)~\cite{BryantBdd} to represent the routing policy for each interface. BDDs can compactly represent Boolean functions and are a canonical representation for such functions. \hl{Memoization combined with uniqueness of the representation means that two BDDs are semantically-equivalent iff their pointers are the same.} This turns checking equivalence of any two transfer functions into an $O(1)$ operation after their BDDs are constructed.

As an example, consider the BGP routing policy in Figure~\ref{fig:bdd-example}. The policy checks if either the 65001:1 or 65001:2 community is attached to an inbound route advertisement. If so, it adds the 65001:3 community and updates the local preference to 350. Each node in the BDD represents a boolean variable used to represent state in the advertisement. Primed variables represent output values after applying updates to the advertisement. A solid arrow means the value is true, while a dashed arrow means the value is false. There are two leaf values: 0 and 1 which represent false and true, respectively. Any path from the BDD root to 1 represents a valid input-output relation. If $c_1$, the variable representing community 65001:1 is true, then the resulting advertisement will have $c_3'$ true (65001:3 attached), and will have a local preference for the 32 bit value representation of 350.

\subsection{The Algorithm}

Algorithm~\ref{alg:alg1} lists the steps used to compute the abstraction function $f$ given graph ($G$) and a collection of BDDs ($bdds$). The first step is to specialize the $bdds$ to the particular destination $G.d$ (line 2). We use a union-split-find data structure to maintain a collection of disjoint sets of concrete nodes that represent the abstract nodes in the network. One of the first steps is to split the collection of sets so that $G.d$ becomes its own abstract node (line 4) and every other concrete node remains as a single other abstract node. Next, it repeatedly tries to refine the abstraction while it is not a \effective abstraction. The algorithm iterates over each current abstract node (set of concrete nodes). If the abstract node is already fully concrete (line 8), then it continues, otherwise it refines the abstraction. $\mathrm{Refine}$ iterates over each concrete node $u$ in the abstract node $\wh{u}$ and each edge from $u$ to $v$, and builds a map from $u$ to a set of pairs of the BDD policy along edge $(u,v)$ and the neighboring node (line 20)---either the concrete neighbor (for \aaabs) or the abstract neighbor (for \aeabs). Finally, we group entries of the map ($us$) by those values that have the same pairs of policies and neighbors, and then refine the abstraction by these groups (line 22). This step ensures that groups of devices that have different transfer functions or policies to different neighbors are separated in the next iteration of the algorithm.

\begin{figure}[t]
  \includegraphics[scale=.214]{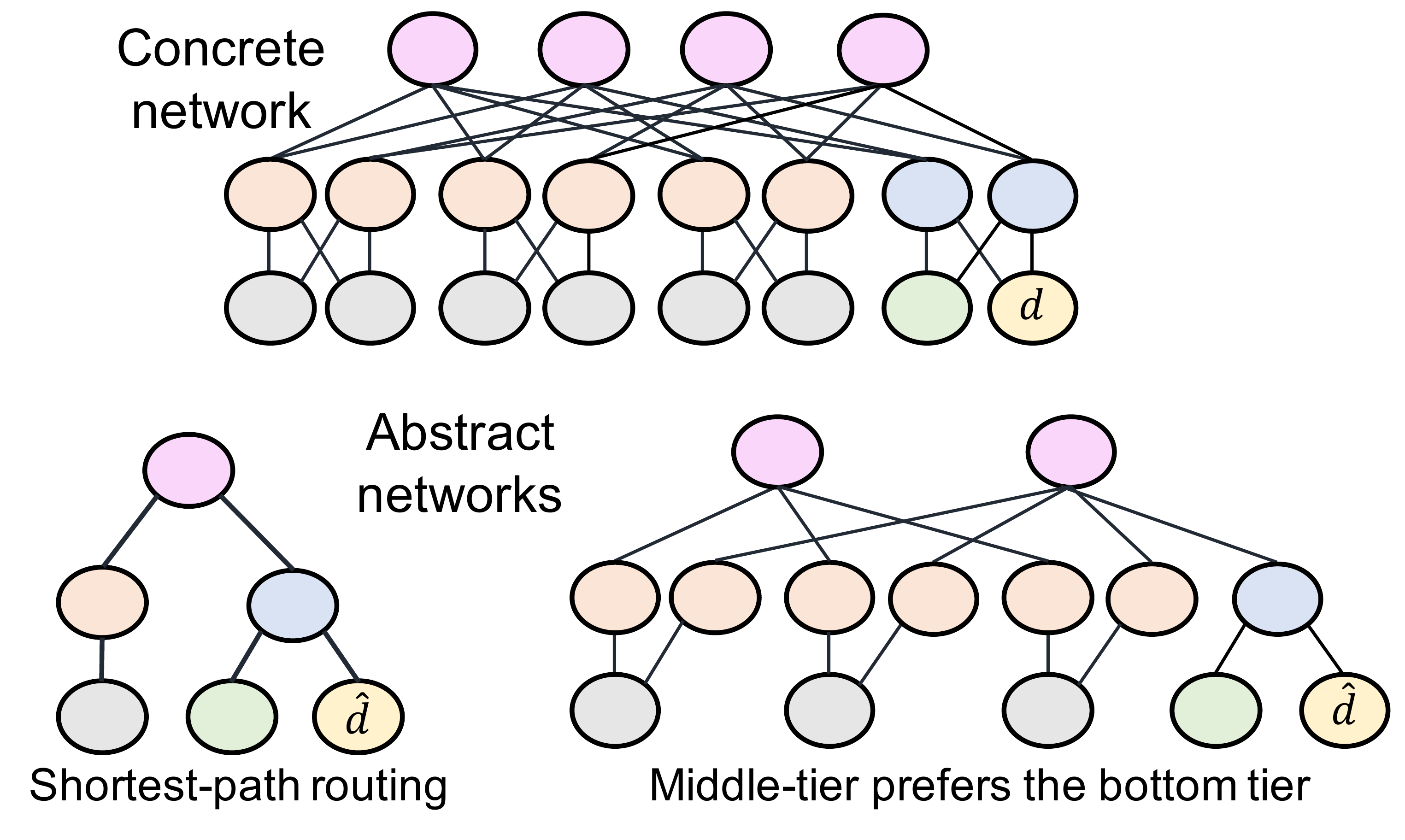}
  \vspace{-1em}
  \caption{Abstractions for a network running BGP on a fattree topology using different policies.}
  \label{fig:example-abstractions}
\end{figure}

Figure~\ref{fig:example-abstractions} shows
the output of the algorithm on a BGP-based fattree network with two different routing policies. In one case, the network uses shortest (AS) path routing, and in the second case, the middle-tier of routers prefer to route via the bottom tier.  The abstract network is bigger in the second case to capture the greater number of possible forwarding behaviors of the middle-tier routers.

%
%
%
%

\begin{table*}[t!]
  \begin{footnotesize}
  \begin{tabular}{|c|c|c|c|c|c|c|c|c|} \hline
    \textbf{Topology} &
      \textbf{Nodes / Edges} &
      \multicolumn{3}{|c|}{\textbf{Abs. Nodes / Edges}} &
      \textbf{Compression ratio} &
      \textbf{Num ECs} &
      \textbf{BDD time} & 
      \textbf{Compression time (per EC)} \\ \hline
    \multicolumn{9}{c}{(a) Synthetic networks} \\ \hline
    \multirow{3}{*}{Fattree} 
    & 180 / 2124 & \multicolumn{3}{|c|}{6 / 5} & 30$\times$ / 424.8$\times$ & 72 & 0.36 & 0.09  \\  
    & 500 / 9100 & \multicolumn{3}{|c|}{6 / 5} & 83.33$\times$ / 1820$\times$ & 200 & 1.29 & 0.26 \\
    & 1125 / 29475 & \multicolumn{3}{|c|}{6 / 5} & 187.5$\times$ / 5895$\times$ & 450 & 7.87 & 0.75 \\ \hline 
  
    \multirow{3}{*}{Ring} 
    & 100 / 100 & \multicolumn{3}{|c|}{51 / 50} & 1.96$\times$ / 2$\times$ & 100 & 0.14 & 0.08 \\
    & 500 / 500 & \multicolumn{3}{|c|}{251 / 250} & 1.99$\times$ / 2$\times$ & 500 & 0.33 & 2.29 \\ 
    & 1000 / 1000 & \multicolumn{3}{|c|}{501 / 500} & 2$\times$ / 2$\times$ & 1000 & 0.34 & 12.26 \\ \hline 
  
    \multirow{3}{*}{Full Mesh} 
    & 50 / 1225 & \multicolumn{3}{|c|}{2 / 1} & 25$\times$ / 1225$\times$ & 50 & 0.18 & 0.07 \\
    & 150 / 4950 & \multicolumn{3}{|c|}{2 / 1} & 75$\times$ / 4950$\times$ & 150 & 1.11 & 0.34 \\ 
    & 250 / 31125 & \multicolumn{3}{|c|}{2 / 1} & 125$\times$ / 31125$\times$ & 250 & 3.31 & 5.48 \\ \hline 
  
    \multicolumn{9}{c}{(b) Real networks} \\ \hline
    Data center & 197 / 16091 & \multicolumn{3}{|c|}{30.2 $\pm$ 2.2 / 143.6 $\pm$ 18.6} & 6.6$\times$ / 112$\times$ & 1269 & 132.28 & 15.51  \\ \hline
    WAN & 1086 / 5430 & \multicolumn{3}{|c|}{209.4 $\pm$ 36.5 / 759.4 $\pm$ 129.2} & 5.2$\times$ / 7.2$\times$ & 845 & 11.35 & 1.83  \\ \hline
    
  \end{tabular}
  \vspace{1.8em}
  \caption{Compression results for synthetic and real networks. All times are in seconds.}
  \label{fig:evaluation}
  \end{footnotesize}
\end{table*}

\section{Practical Extensions}
\label{sec:practical-extensions}


\para{Multiple Protocols}
Although the stable routing problem is framed in terms of the behavior of a particular protocol, devices in practice often run multiple protocols at once. \hl{One can build a new SRP to model these interactions. For example, if a network runs both OSPF and eBGP, then the SRP could use attributes of the form} $\hlmath{A = A_{BGP} \times A_{OSPF} \times A_{RIB}}$. \hl{That is, track both OSPF and BGP, as well as} $\hlmath{A_{RIB}}$, \hl{which represents the main RIB that carries the best route (based on administrative distance) between the various protocols and records what protocol was chosen. Following ideas from Batfish}~\cite{batfish},
\hl{we model route redistribution, where routes from one protocol are injected into another, via the transfer function. For instance, if OSPF routes are redistributed into BGP, then BGP will allow routes from} $\hlmath{A_{RIB}}$ \hl{even when they are from OSPF.}
 
\para{Access Control Lists}
While ACLs do not affect control plane routing information, they can prevent traffic from being forwarded out an interface. For this reason, we conservatively consider the ACL to be part of the transfer function, which gets captured in the BDD, so that nodes will only be abstracted together if they have the same ACLs with respect to destination $d$. This ensures that the fwd-equivalence property will remain valid.

\para{iBGP}
iBGP is a complicated protocol that recursively routes packets for eBGP by communicating them over an IGP path. If there is a valid abstraction for both the IGP and for eBGP, and there is no ACL in the network that blocks iBGP loopback addresses, then multiple iBGP neighbors can be compressed together. \hl{This is because (1) both iBGP neighbors will be sent the same eBGP routes from neighbors, (2) these advertisements will have the same IGP cost metric (since they must be symmetric with respect to the IGP as well), and (3) although the iBGP neighbors may have an edge between them, potentially violating the self-loop-free requirement, this edge is never used since iBGP does not re-advertise routes learned over iBGP to other iBGP neighbors.}

%
%
%
%

\section{Implementation}
\label{sec:implementation}


We implemented our network abstraction algorithm in a tool called \sysname. It uses the Batfish~\cite{batfish} network analysis framework to convert network configurations into a vendor-independent intermediate representation. \sysname operates over this vendor-independent format to create a network abstraction in the form of a smaller, simpler collection of vendor-independent configurations. 
Tools built using this framework, such as Batfish and Minesweeper, can then work with the smaller configurations to speed up their analysis. 

We use the Javabdd~\cite{javabdd} library to encode router-level import and export filters, as well as access control lists (ACLs) as BDDs.
Because \sysname creates abstract networks per destination EC, and such ECs are disjoint,our implementation is able to generate abstract networks and check their properties in parallel. We only generate abstract networks for destination ECs that are relevant for a query. For example, checking port-to-port reachability would typically only require generating a single abstract network for one EC.

%


%
%
%
%

\begin{figure*}
  \subfloat[Fattree]{%
    \includegraphics[scale=.24]{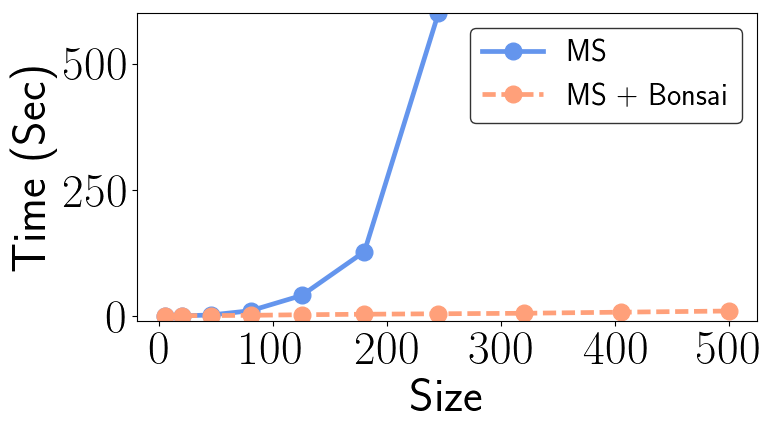}
  }
  \hspace{2em}
  \subfloat[Full Mesh]{%
    \includegraphics[scale=.24]{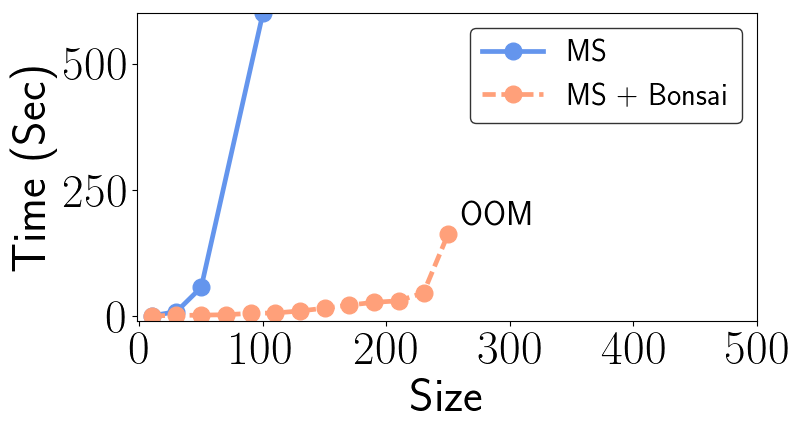}
  }
  \hspace{2em}
  \subfloat[Ring]{%
    \includegraphics[scale=.24]{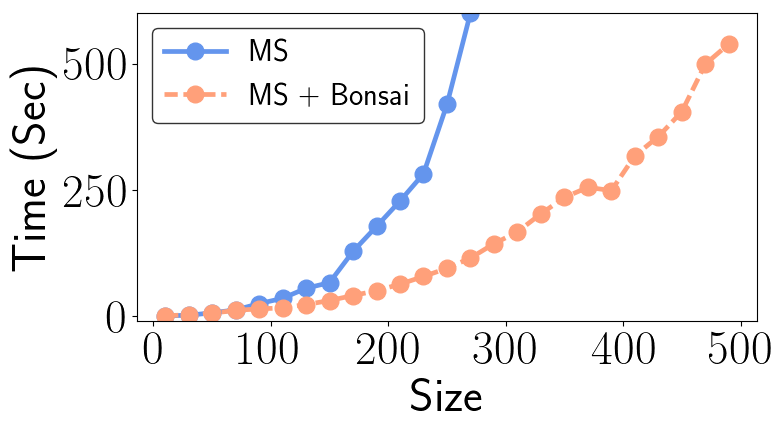}
  }

  \vspace{-.4em}
  \caption{Minesweeper (MS) verification time with and without abstraction for an all-pairs reachability query.}
  \label{fig:verification-time-sp}
\end{figure*}

\section{Evaluation}
\label{sec:evaluation}

We evaluate \sysname using a collection of synthetic and real networks. We aim to answer the following questions: (i) can \sysname scale to large networks? (ii) can its algorithm effectively compress networks? and (iii) can the abstract, compressed networks be speed network analysis?

\para{Networks studied} 
We study three types of synthetic network topologies: Fattree~\cite{fattree}, Ring, and Full-mesh. Each such network uses eBGP to perform shortest path routing along with destination-based prefix filters to each destination. These networks are highly symmetric by design and are used to characterize compression as a function of network topology and size. For each topology type, we scale the size and measure the effectiveness and cost of compression. 

While the synthetic networks focus on the effect of topology on compression, in practice, most networks do not have perfect symmetry. For this reason, we study operational networks of two different corporations.  The first is a datacenter network with 197 routers organized into multiple clusters, each with a Clos-like topology (rather than a single, large Clos-like topology). The network primarily uses eBGP and static routing, with each router running as its own AS using BGP private AS numbers. It also makes extensive use of route filters, ACLs, and BGP communities. All together, it has over 540,000 lines of configuration. Although there are less than 200 routers in the network, there are over 16,000 physical and virtual interfaces in the network.

The second operational network is a wide-area network (WAN) with 1086 devices, which are a mix of routers and switches. The network uses a eBGP, iBGP, OSPF, and static routing, and consists of over 600,000 lines of configuration.

\para{Synthetic network results}
Table~\ref{fig:evaluation}(a) shows the results of running \sysname on the synthetic networks.
All experiments were run on an a 8-core Macbook Pro with an Intel i7 processor and 8GB of RAM. For each synthetic network, \sysname is able to compress the network quickly. For instance, the largest Fattree topology with 1125 nodes takes around 7.9 seconds to build the BDD data structures and an average of of .75 seconds per EC to compute the abstract network for the 450 ECs. Because equivalence classes are processed in parallel, it takes under a minute to abstract this network. The compressed network size computed is 6 nodes.

For the Fattree and Full-mesh topologies, the compressed network size stays constant as the concrete network grows.
For the ring topology, the compressed network size does grow with the size of the network, and in particular, grows with the diameter of the network. This is necessary since the abstraction must preserve path length. Computing an abstraction for the ring topologies is more expensive because the compression algorithm is only able to split out a single new abstract role with each iteration. 

\sysname's compression has a large effect on network analysis time. Figure~\ref{fig:verification-time-sp} shows the total verification time to check an all-pairs reachability query compared to topology size for each type of synthetic network using Minesweeper~\cite{minesweeper}. We use a timeout of 10 minutes.  The verification time for abstract networks includes the time used to partition the network, build the BDDs, and compute the compressed network. In all cases, abstraction significantly speeds up verification even when taking into account the time to run \sysname. Abstracting the Full-mesh topology ran out of memory beyond a certain point, due to the density of the topology.

\para{Real network results}
For both networks, we first computed the BDDs and see how many devices have identical transfer functions from their configurations. In the datacenter network, we initially found that there were 112 unique "roles" (set of policies) among the 197 routers. However, many of these differences could be attributed to BGP community values that were attached to routers, but then never matched on in any configuration file. To account for these differences, we use the abstraction function for BGP: $h(\lp, \tags, \aspath) = (\lp, \tags - \{\mathrm{unused}\}, f(\aspath))$, which ignores differences from such irrelevant tags. With this abstraction function, we find that there are only 26 unique "roles" among the 197 routers. Further, most of the differences are due to differences in static routes in the configurations. Without static routes, there would only be 8 unique roles. Table~\ref{fig:evaluation} (b) shows the compression results from this network. It takes just over 2 minutes to compute the BDDs and roughly 15 seconds on average to compute a good abstraction per EC. This time is mainly due to the huge number of virtual interfaces. The average compressed network size is around 30 nodes (a 6.6x reduction), and around 132 edges (a 112x reduction).

For the WAN, we found 137 unique "roles" among the 1086 devices. Many of the differences are from neighbor-specific, prefix-based filters and ACLs. It takes around 11 seconds to compute the BDDs for the network, and under 2 seconds per EC to compute a good abstraction. The average compressed size achieves a 5.2x reduction in the number of nodes and a 7.4x reduction in the number of edges.

Finally, to test the effectiveness of \sysname at facilitating scalable analysis of real networks, we run a reachability query between two devices in Batfish, both with and without abstraction. Batfish first simulates the control plane to produce the data plane and then uses NoD~\cite{nod} to compute all possible packets that can traverse between source and destination nodes. With \sysname, it takes 77 seconds to complete the query. Without it, the query did not complete and gave an out-of-memory error after running for over an hour.

%
%
%
%

\section{Related Work}
\label{sec:related}

 
\para{Network verification}
%
\hl{The field of network verification may be split into data plane verification}~\cite{hsa, anteater,nod,netplumber,veriflow,atomic,ddnf,delta-net} and control plane verification~\cite{rcc,batfish,arc,era,bagpipe,minesweeper},\hl{ with our work sitting in the latter camp.  However, Bonsai is orthogonal to,
  and synergistic with, most of this reseach as it compresses networks and leaves the analysis to
  other tools, which typically operate much more quickly over the compressed network.
  Bonsai works because large networks typically contain symmetries, an observation made and exploited Plotkin}~\ETAL~\cite{surgery}, though Plotkin \ETAL focus on data plane
  properties whereas we focus on control plane properties.   The only
  other control plane compression work we are aware focuses on compressing
  BGP networks using local rewrites to preserve convergence properties~\cite{bgp-compression}. \hl{In contrast, we introduce the SRP model to compress networks running a wide-variety of protocols using both local and some non-local (BGP splitting) rewrites.  We aim to preserve forwarding properties so network
  administrators can test for reachability, access control and other
  path-based properties rather than convergence.}

\para{Control plane models} A formal model of network control plane planes lies at the heart of our work. Many prior works have developed such models to describe formally the computation of routing protocols, their safety criteria, or to generalize their computation~\cite{stable-paths,routingalgebra,metarouting}. Our model, SRP, is inspired by Griffin \ETAL's stable paths problem (SPP) which described control plane solutions computed by path vector protocols~\cite{stable-paths}. While both models describe stable solutions, SRP formalizes device-level processing of routing information instead of end-to-end paths. This difference allows it to capture a broader range of control plane features. 

\hl{SRPs are similar to routing algebras}~\cite{routingalgebra, metarouting}, \hl{though we have simplified our presentation slightly by allowing graph edges to stand in for the labels used in routing algebras. However, the more significant difference is that while routing algebras have been used to study convergence properties, which are independent of network topology, we study topology-dependent properties such as reachability, and developed compression algorithms that preserve such properties.}

 
\para{Abstractions in verification} Conservative abstractions are the mainstay of program verification in various forms such as loop invariants~\cite{Floyd67, Hoare69}, abstract interpretation~\cite{CC77}, and counterexample guided abstraction refinement~\cite{satabs, ClarkeCegar00, slam}. These abstractions enable sound analysis for verification problems that are often undecidable or intractable. Tighter abstractions based on symmetry and bisimulations have also been used successfully to scale model checking~\cite{EmersonSymm93,ClarkeSymm93}. 
We build on these foundations to seek useful abstractions for compressing networks that preserve \cpequivalence.

%
%
%
%

\section{Conclusion}
\label{sec:conclusions}

Recently, researchers have made great progress in control plane analysis, using a variety of techniques ranging from simulation to verification. But the scale and complexity of real networks often renders such techniques computationally expensive or even intractable. To accelerate analysis, our \sysname tool automatically compresses a network and its configurations by eliminating any structural symmetries. \sysname is based on a theory of control plane equivalence of two networks and an efficient compression algorithm. We show that it scales well and effectively compresses real networks.

\para{Acknowledgements}
We would like to thank the ACM SIGCOMM reviewers and our shepherd
Laurent Vanbever, whose feedback helped improve this paper.
This work was supported in part by NSF Grants 1703493 and 1525936, and
gifts from Cisco and Facebook. Any opinions, findings, and conclusions
expressed herein are those of the authors and do not necessarily
reflect those of the NSF, Cisco or Facebook.

%
%
%
%

\balance

\bibliographystyle{abbrv}
\bibliography{references}

%
%
%
%

\appendix
\onecolumn
 

\section{Proof of \CPequivalence}

Here we give the full proof of \cpequivalence from Section~\ref{ssec:fwd-equivalent}. The proof requires additional lemmas and definitions not introduced in Section~\ref{ssec:fwd-equivalent}. First, we make an observation about attribute equality ($\approx$).

\vspace{1em}
\begin{thm} \label{thm:rank-preservation}
  Given an \effective abstraction, $\forall a,b.~ a \approx b \iff h(a) \approx h(b)$
    \begin{proof}
    Immediate from rank-equivalence. Suppose $a \approx b$. Then $a \not\prec b \wedge b \not\prec a$. From rank-equivalence, this means that $h(a) \not\prec h(b) \wedge h(b) \not\prec h(a)$, and thus $h(a) \approx h(b)$. The reverse holds by the same reasoning.
    \end{proof}
\end{thm}


Next, we define \emph{choice-equivalence}, which states that nodes in the abstract and concrete networks receive similar types of choices from similar neighbors:

\vspace{1em}
\begin{defn}
We say that an abstraction $(f,h)$ is \emph{choice-equivalent} if the following holds:
\[ \begin{array}{l}
 1.~~ \forall e,a.~ (e,a) \in \choices(u) \implies (f(e), h(a)) \in \wh{\choices}(f(u)) \\
 2.~~ \forall E,A.~ (E,A) \in \wh{\choices}(f(u)) \implies \forall e \mapsto E,~ \exists a.~ a \mapsto A ~\wedge~ (e,a) \in \choices(u) \\
\end{array} \]
\end{defn}

\vspace{1em}
\begin{thm} \label{thm:pref-equiv}
If we have a self-loop-free $SRP$ and $\wh{SRP}$, and a \effective abstraction that is choice-equivalent, then the abstraction is label-equivalent.

  \begin{proof} 
  Looking at the definition of $\lab$, there are 3 cases to consider. First we observe that if $v = d$, then $\lab(d) = a_{d}$. It follows that $\wh{\lab}(f(d)) = \wh{\lab}(\wh{d}) = \wh{a}_{d} = h(a_{d}) = h(\lab(d))$. In the second case, using choice-equivalence and \aeabs, we can see that $\attrs(v) = \emptyset \iff \wh{\attrs}(f(v)) = \emptyset$. Thus, $h(\lab(v)) = h(\bot) = \bot = \wh{\lab}(f(v))$. For the final case with $\attrs(v) \neq \emptyset$, we show the implications separately.
  
  \vspace{.8em}
  \noindent \textbf{Case ($\Rightarrow$)}
  Assume $\lab(v) = a$. By the definition of $\lab$, we know that $a \in \attrs(v)$ and is minimal by $\prec$. We know that there is some edge $e$ such that $(e,a) \in \choices(v)$. Consider all concrete edges $(e',a') \in \choices(v)$. From choice-equivalence, we know that $(f(e'),h(a')) \in \wh{\choices}(f(v))$ for each such pair. From rank-equivalence, we know $(f(e),h(a)) \in \wh{\choices}(f(v))$ and is minimal by $\wprec$. By the definition of $\lab$, we then know that $\wh{\lab}(f(v)) = A = h(a)$. By transitivity, $\wh{\lab}(f(v)) = h(\lab(v))$

  \vspace{.8em}
  \noindent \textbf{Case ($\Leftarrow$)}
  Assume $\wh{\lab}(f(v)) = A$. We know that $A \in \wh{\attrs}(f(v))$ and is minimal by $\wprec$. Assume $(E,A) \in \wh{choices}(f(v))$. Consider all such $(E',A') \in \wh{\choices}(f(v))$. From choice-equivalence, we know that for any concrete edge $e \mapsto E$, there exists $a$ such that $h(a) = A$ and $(e,a) \in choices(v)$. Therefore, $a \in \attrs(v)$. Let us consider a smallest such $a$ in terms of $\prec$. From rank-equivalence, we know that $a$ is smaller than any $(e',a')$ where $f(e') \neq f(e)$ since $A$ was the smallest such value in $\wh{choices}$. Therefore, $a \in attrs(v)$ and is minimal by $\prec$. Finally, we obtain that the labeling can be $a$ ($\lab(v) = a$). It follows from transitivity that $h(\lab(v)) = \wh{\lab}(f(v))$.
  \end{proof}
\end{thm}

\vspace{1em}
\begin{thm} \label{thm:forward-equiv}
If we have a self-loop-free $SRP$ and $\wh{SRP}$ and a \effective abstraction that is choice-equivalent, then the abstraction is fwd-equivalent.

\begin{proof} \text{ }
  From Theorem~\ref{thm:pref-equiv}, we know that we have label-equivalence. 
  
  \vspace{.8em}
  \noindent \textbf{Case 1}
  We assume $e = (u,v) \in \fwd(u)$ and need to show that $f(e) \in \wfwd(f(u))$. By the definition of $\fwd$, we know that $\exists a. (e,a) \in \choices(u) \wedge a \approx \lab(u)$. From choice-equivalence, this means that 
  $$(f(e),h(a)) \in \wh{\choices}(f(u))$$
  Thus, because we have choice-equivalence, we have label-equivalence. Recall from label equivalence: $h(\lab(u)) = \wlab(f(u))$.
  From Theorem~\ref{thm:rank-preservation}, we have $a \approx \lab(u)$ so $h(a) \approx h(\lab(u))$ and thus $h(a) \approx \wlab(f(u))$. Then, by the definition of $\wfwd$:
  $$f(e) \in \wh{\fwd}(f(u))$$

  \vspace{.8em}
  \noindent 
  \textbf{Case 2}
  We will assume $(\wh{u},\wh{v}) \in \wh{\fwd}(\wh{u})$ and show that for all concrete nodes $u \mapsto \wh{u}$, there exists a $v \mapsto \wh{v}$ such that $(u,v) \in \fwd(u)$. By the definition of $\fwd$, we know that: $\exists A.~ ((\wh{u},\wh{v}),A) \in \wh{\choices}(\wh{u}) \wedge \wh{\lab}(\wh{u}) \approx A$.
  From choice-equivalence, this means:
  $$\forall e \mapsto E,~ \exists a.~ h(a) = A \wedge (e,a) \in \choices(u) \wedge \wh{\lab}(\wh{u}) \approx A$$
  Consider any such $e = (u,v)$ where $f(e) = E$. Rewriting slightly, we get:
  $$\exists a.~ (e,a) \in \choices(u) \wedge \wh{\lab}(f(u)) \approx h(a)$$

  Once again, from Theorem~\ref{thm:rank-preservation} and transfer-equivalence, we know that $\wlab(f(u)) = h(\lab(u))$ and so $h(a) \approx h(\lab(u))$, and therefore: $a \approx \lab(u)$. Finally, from the definition of $\fwd$ we have 
  $$e \in \fwd(u)$$.

\end{proof}
\end{thm}

\begin{thm} \label{thm:dag}
  The forwarding behavior for any solution $\lab$ to a well-formed, loop-free SRP will form a DAG rooted at the destination $d$.
  \begin{proof}
    We know that the solution is loop-free so the result must not have cycles. Also, there can only be one root for the DAG ($d$) because if there were another $d'$, then $\lab(d') = \bot$, otherwise $d'$ would forward to some neighbor. However, because the SRP is non-spontaneous, this can not happen.
  \end{proof}
\end{thm}

\vspace{1em}
\begin{thm} \label{thm:correctness-local}
  A well-formed, loop-free $SRP$ and its \effective abstraction $\wh{SRP}$ are label- and fwd-equivalent. That is, for any $\lab$ there exists label and fwd-equivalent $\wlab$ and vice-versa.

  \begin{proof} 
    It suffices to first show choice-equivalence. We then get label-equivalence for free from Theorem~\ref{thm:pref-equiv}, and then that $SRP$ and $\wh{SRP}$ are fwd-equivalent from Theorem~\ref{thm:forward-equiv}.

    Because we know the SRP is loop-free and non-spontaneous, we know that any stable solution $\lab$ (and $\wlab$) must form a rooted DAG at the destination $d$ (Theorem~\ref{thm:dag}). We start by showing a slightly strengthened inductive hypothesis: with the choice-equivalence property above for the subgraph corresponding to the actual forwarding edges in the provided concrete (or abstract) solution $\lab$ (or $\wlab$). That is, given a concrete solution $\lab$ we will only consider edges $e = (u,v)$ where $e$ goes from level $k+1$ to level $k$ in the DAG, and similarly for the abstract network, we will only consider the corresponding edges $f(e)$. Symmetrically, for the other direction, we will only consider abstract edges $\hat{e} = (\wh{u}, \wh{v})$ going from level $k+1$ to level $k$ and only the edges $e$ where $f(e) = \wh{e}$ for the concrete network. For both directions, we will show label-equivalence ($h(\lab(v)) = \wh{\lab}(f(v))$) holds at each node. We show each direction of the stronger implication separately, using induction on the level of the DAG.
    
    \vspace{.8em}
    \noindent \textbf{Base case (for $\Rightarrow$ and $\Leftarrow$):}
    For the base case, from the definition of $\lab$, we know that $\lab(d) = a_{d}$ and $\wh{\lab}(\wh{d}) = \wh{a}_{d}$. 
    From dest-equivalence, we know that $f(d) = \wh{d}$, so:
    $$\wh{\lab}(f(d)) = \wh{\lab}(\wh{d}) = \wh{a}_{d} = h(a_{d}) = h(\lab(d))$$
    
    Since there are no edges $e$ going to a lower level in the DAG (than the root) in either the concrete or abstract, we are done.
    
    \vspace{.8em}
    \noindent \textbf{Inductive case ($\Rightarrow$)}
    We are given $\lab$ and show that there exists a $\wlab$ for the subgraph we have induced that has label-equivalence. Consider an arbitrary node $u$ at depth $k$. Now, suppose $(e,a) \in \choices(u)$ and $e = (u,v)$. We know that $v$ appears at level $k-1$ in the DAG. We also know that $a = \transfer(e, \lab(v)) \neq \bot$. Since $a \neq \bot$, from drop-equiv, we know that $h(a) \neq bot$. By the IH with label-equivalence, we know that $\wh{\lab}(f(v)) = h(\lab(v))$. From transfer-equivalence, we know that 
    $$\wh{\transfer}(f(e), h(\lab(v))) = h(\transfer(e, \lab(v))) = h(a) \neq \bot$$
    By transitivity and label-equivalence (IH) then, we know:
    $$\wh{\transfer}(f(e), \wh{\lab}(f(v))) = h(a)$$
    By the definition of $\choices$, it follows that
    $$(f(e),h(a)) \in \wh{\choices}(f(u))$$
    Hence, we have choice-equivalence. This means that the set of choices available at $f(u)$ from $f(v)$ is ``the same'' as the set of choices available at $u$ from $v$. Since we have choice-equivalence, it follows that we have label-equivalence and fwd-equivalence for the subgraph under consideration. 
    
    \vspace{.8em}
    \noindent \textbf{Inductive case ($\Leftarrow$)}
    We are given $\wlab$ and show that there exists a $\lab$ for the induced subgraph that has label-equivalence. Consider an arbitrary node $\wh{u}$ at depth $k$ of the abstract subgraph. Now, suppose $(E,A) \in \wh{\choices}(\wh{u})$ and $E = (\wh{u},\wh{v})$. From the \aeabs and the fact that $f$ is onto, we know there must be at least some $e$ such that $f(e) = E = (f(u),f(v))$ (otherwise $E$ could not have been an abstract edge). Consider an arbitrary such $e = (u,v)$. We know that $\wh{v}$ appears at level $k-1$ in the DAG (and so does $v$ by construction). We also know that $A = \wh{\transfer}(f(e), \wh{\lab}(f(v)) \neq \bot$. From drop-equiv, we know that any $a$ where $h(a) = A$ and $A \neq \bot$ has $a \neq \bot$. As before, we observe by the IH that $\wh{\lab}(f(v)) = h(\lab(v))$. And so:
    $$A = \wh{\transfer}(f(e), h(\lab(v))) = h(\transfer(e, \lab(v))) \neq \bot$$ 
    Let $a$ stand for $\transfer(e, \lab(v))$. Then $a = \transfer(e, \lab(v))$ and $h(a) = A$. By the definition of $\choices$, it follows that: 
    $$\exists a.~ h(a) = A \wedge (e,a) \in \choices(u)$$
    Because we showed choice equivalence for any such edge $e$ from any node $u \mapsto \wh{u}$, we have choice-equivalence. This implies that we also have label- and fwd-equivalence.

    \vspace{.8em}
    \noindent \textbf{Other edges}
    All that remains is to show that edges going to a equal or higher level of the DAG do not change the existing solution. Suppose we were given the concrete network. Consider such an edge $\wh{e} = (\wh{u}, \wh{v})$. For this edge to affect the current solution $\wlab$, it must be the case that for some $\wh{e}'$ and $\wh{v}'$:
    $$\wtransfer(\wh{e}, \wlab(\wh{v})) ~\wprec~ \wlab(\wh{u}) = \wtransfer(\wh{e}', \wlab(\wh{v}'))$$
    
    Rewriting slightly:
    $$\wtransfer(\wh{e}, h(\lab(v))) ~\wprec~ \wtransfer(\wh{e}', h(\lab(v'))))$$

    From transfer equivalence:
    $$h(\transfer(e, \lab(v))) \prec h(\transfer(e', \lab(v')))$$

    From rank-equivalence:
    $$\transfer(e, \lab(v)) \prec \transfer(e', \lab(v'))$$

    Given the definition of $\lab$, this leads to a contradiction with the fact that the concrete solution was indeed stable. In particular, node $u$ has a better option through $v'$ over $v$, and hence, the labelling is incorrect. We can conclude then, that here can be no such better option in the abstract network. The argument is symmetric for the other direction.
  \end{proof}
\end{thm}

Using Theorem~\ref{thm:correctness-local}, we may conclude that any \effective{} abstractions of common protocols, which produce loop-free routing, are \cpequivalent. Now we show that static routing, which is not necessarily loop-free, also has this property.

\vspace{2em}
\begin{thm} \label{thm:correctness-static}
  Given self-loop-free $SRP$ and $\wh{SRP}$ for static routing with an
  \effective{} abstraction, then it is fwd-equivalent.
   
  \begin{proof}
    Because the labeling at each node does not depend on the labeling at other nodes,
    the proof is direct. 
    As before, we show choice-equivalence, then rely on~\ref{thm:forward-equiv} to derive \cpequivalence.

    \vspace{.8em}
    \noindent \textbf{Case ($\Rightarrow$)}
    Assume $e = (u,v)$. We have $(e,a) \in \choices(u)$. By unfolding the definition of $\choices$, we know that $a = \transfer(e, \lab(v))$. By transfer equivalence, we know that 
    $$h(a) = h(\transfer(e, \lab(v))) = \wh{\transfer}(f(e), h(\lab(v)))$$
    There are now 2 cases. Suppose $a = 1$. Then $h(a) = 1$, so 
    $$\wh{\transfer}(f(e), h(\lab(v))) = 1$$
    The definition of $\transfer$ does not depend on the attribute for static routes, we know that :
    $$\wh{\transfer}(f(e), \wh{\lab}(v)) = 1$$
    It follows that 
    $(f(e),1) \in \wchoices(f(u))$
    
    \noindent The case for $a = 0$, is symmetric, with $(f(e),0) \in \wchoices(f(u))$.

    \vspace{.8em}
    \noindent \textbf{Case ($\Leftarrow$)}
    Suppose $((\wh{u},\wh{v}), A) \in \choices(\wh{u})$. We need to show that for any edge $e \mapsto E$, there exists an $a$ such that $(e,a) \in \choices(u)$ and $h(a) = A$. Let us choose $a = A$ for static routes. Clearly $h(a) = A$ since $h$ is the identity. Consider arbitrary edge $e \mapsto E$. We have:
    $$ \wh{\transfer}(f(e), \wh{\lab}(f(v)))) = A = h(a) = a$$
    Again, since the definition of $\wh{\transfer}$ does not depend on the neighbor attribute, we can replace it with any value. In particular, this is the same as:
    $$a = \wh{\transfer}(f(e), h(\lab(v)))$$
    From transfer-equivalence and transitivity we know that:
    $$a = \transfer(e,\lab(v))$$
    Finally, from the definition of $\choices$:
    $\exists a.~ h(a) = A \wedge (e,a) \in \choices(u)$
  \end{proof}
\end{thm}

\begin{corollary} \label{thm:all-protocol-equiv}
  Suppose we have a self-loop-free $SRP$ and $\wh{SRP}$ for RIP, OSPF,  static routing, or BGP (without loop prevention), related by \effective{} abstraction $(f,h)$. There is a solution $\lab$, where each node $u_1 \mapsto \wh{u_1}$ forwards along label path $s = \lab(u_1) \ldots \lab(u_k)$ to some node $u_k \mapsto \wh{u_k}$ \ifft there is a solution $\wlab$ that forwards along the label path $\wh{s} =  \lab(\wh{u_1}) \ldots \lab(\wh{u_k})$ and $h(s) = \wh{s}$.

  \begin{proof} We show each direction separately.

    \vspace{.8em}
    \noindent \textbf{Case ($\Rightarrow$)}
    Suppose $\lab$ is a solution for $SRP$.      Given any two nodes $u$ and $v$ where $u$ can reach $v$, there exists a path $p = u,w_1,\ldots,w_k,v$ where $(u,w_1) \in \fwd(u)$ and $(w_i,w_{i+1}) \in \fwd(w_i)$ and $(w_k,v) \in \fwd(w_k)$. Because $\lab$ and $\wh{\lab}$ are fwd-equivalent, we know that $(f(u),f(w_1)) \in \wfwd(f(u))$ and so on. 
    Therefore, there is an abstract path in $\wh{\lab}$ where $f(u)$ can reach $f(v)$ where the path has the form $f(u), f(w_1), \ldots, f(w_k), f(v)$. 
    The labels of the concrete path are $s = \lab(u), \lab(w_1), \ldots, \lab(w_k), \lab(v)$. Similarly, the abstract path has labels $\wlab(f(u)), \ldots, \wlab(f(v))$. It follows from label-equivalence that 
    $\wlab(f(u)), \ldots, \wlab(f(v)) = h(\lab(u)), \ldots, h(\lab(v))$. Finally, the definition of $h$ gives us: 
    $h(\lab(u)), \ldots, h(\lab(v)) = h(s)$

    \vspace{.8em}
    \noindent \textbf{Case ($\Leftarrow$)}
    Symmetric to the first case. Suppose $\wh{L}$ is a solution for $\wh{SRP}$. Consider an arbitrary path $\wh{u},\wh{w_1},\ldots,\wh{w_k},\wh{v}$. Then we know $(\wh{u},\wh{w_1}) \in \wh{\fwd}(\wh{u})$ and so on. From the fact that $\lab$ and $\wh{\lab}$ are fwd-equivalent, every node $u$ where $u \mapsto \wh{u}$ will follow some path $(u,w_1) \in fwd(u)$ and so on. Therefore, there will be a concrete path $u, w_1, \ldots, w_k, v$ such that $w_i \mapsto \wh{w_i}$, and $v \mapsto \wh{v}$. The abstract path $\wh{s} = \wlab(\wh{u}), \ldots, \wlab(\wh{v})$. Similarly, the concrete path will have $s = \lab(u), \ldots, \lab(v)$. To show that $\wh{s} = h(s)$, we simply use label-equivalence: $h(s) = h(\lab(u)), \ldots, h(\lab(v)) = \wh{s}$.

  \end{proof}
\end{corollary}

\begin{figure}
  \begin{center}
  \includegraphics[scale=.32]{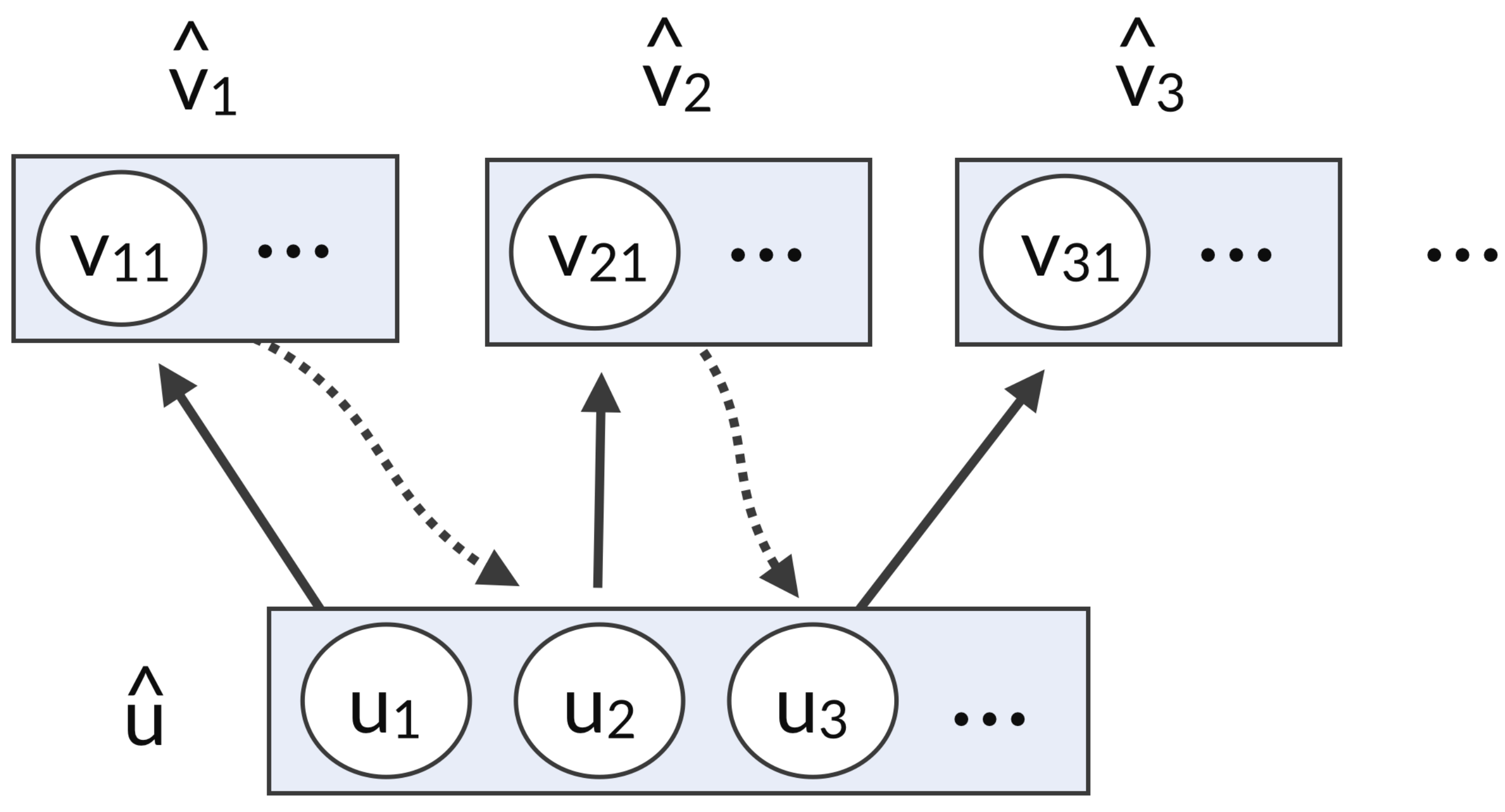}
  \end{center}
  \caption{A stable solution with the maximum number of behaviors.}
  \label{fig:bounded-behaviors}
  \vspace{2em}
\end{figure}

\vspace{1em}
\begin{thm} \label{thm:bounded-behaviors}
  If a well-formed $SRP$ and $\wh{SRP}$ for BGP has an \aaabs and is transfer-approx, then for all solutions $\lab$ to $SRP$, and all abstract nodes $\wh{u} \in \wh{V}$, $\abs{\behaviors_\lab(\wh{u})} \leq \abs{\prefs(\wh{u})}$.

  \begin{proof}
  Because we have rank-equivalence and an \aaabs, the only way two nodes will forward to different neighbors is the transfer functions are different. Otherwise, both nodes would receive the same $\choices$ as in Theorem~\ref{thm:correctness-local} and because of the universal abstraction, they both have an edge to the best such choice and will use this neighbor. 
  Due to relative-transfer-equivalence, the only time this can occur is when two nodes have different transfer functions due to loop prevention.
  
  First we show that there can be $|\prefs(\wh{u})|$ different behaviors. Consider Figure~\ref{fig:bounded-behaviors}. In the figure, $\wh{u}$ has a local preference for $\wh{v}_1$ over $\wh{v}_2$ over $\wh{v}_3$ etc. In this case, $|\prefs(\wh{u})| = 3$. There is a stable solution where $u_1$ forwards to $v_{11}$ since that is the best path. $u_2$ would prefer to use this path, but cannot because it is already on the path, so it cannot consider $v_{11}$ due to its transfer function. Instead, $u_2$ will use the next best choice $v_{21}$. Similarly, $u_3$ would like to use $v_{11}$ or $v_{21}$ but cannot due to loops. Therefore, $u_3$ will forward to $v_{31}$ instead. Because there is a universal abstraction (full mesh), and because we have rank-equivalence and relative-transfer-equivalence, each node has the same choices modulo loops. Such a chain as shown in Figure~\ref{fig:bounded-behaviors} is the only way we can get such different behavior. 
  
  Now we show that there can not be more than $|\prefs(\wh{u})|$ behaviors. The proof is by contradiction. Suppose we have another node $u_4$ and $u_4$ will forward to a different node that each of $u_1$ through $u_3$. $u_4$ can not continue the chain by falling back to the next lowest local preference since all local preferences have been exhausted by $u_1$ through $u_3$. Therefore, $u_4$ will forward to one of the same neighbors as $u_1$ through $u_3$. But this contradicts the assumption. Therefore, there can not be more than $|\prefs(\wh{u})|$ behaviors.

  \end{proof}
\end{thm}

\vspace{1em}
\begin{thm} \label{thm:refinement-equivalence}
  Suppose we have well-formed $SRP$, $\wh{SRP}$, and $\ol{SRP}$ for BGP with an \effective abstraction $(f, h)$. For any solution $\lab$ to $SRP$, there exists a refinement $(f_r, h_r) \sqsubseteq_{(f_s,h_s)} (f, h)$ where $\ol{\lab}$ is a solution to $\ol{SRP}$, and $\lab$ and $\ol{\lab}$ are label- and fwd-equivalent.

  \begin{proof}
    First, we will show a particular refinement.
    From Theorem~\ref{thm:bounded-behaviors}, we know that any solution to $\lab$ can only have $|\prefs(\wh{v})|$ behaviors.
    Let use define $f_r(v) = \ol{v} = \behavior(f(v))(i)$, where this notation means that we pick out the $i$th node in $\ol{V}$ such that $f_s$ maps it to $\wh{v}$.
    We can modify this scheme slightly to ensure that $f_r$ is an \emph{onto} function, if no node would map to the $k$th behavior, then pick an arbitrary node that maps to the $j$th behavior (if there is more than one node that maps to the $j$th behavior), and map it to the $k$th behavior instead.
    This is a valid refinement to $(f,h)$ since $f = f_s \circ f_r$ and $f_r$ is onto.
    
    Since we have a particular solution $\lab$ that is loop-free (since BGP is loop-free), we know all the edges in $SRP$ that are not used due to loops. 
    For example, in Figure~\ref{fig:refinement} in the concrete network (left), the green node would have transfer function $\bot$ from each of the red neighbors below due to loop prevention.
    
    Consider an isomorphic network $G'$, where all such edges are removed (e.g., directed edges from the green to red nodes). Similarly, in the refined network $\ol{G}'$, we would remove the corresponding edges (e.g., the directed edge from the green to red nodes).

    The particular refinement $f_r$ we chose is important because we will still have an \aaabs after removing these edges since each node with such unique behavior rejected the same nodes (due to loops) to accept the worse path. Therefore, removing the abstract edge and concrete edges remains a universal abstraction.

    The same solution $\lab$ is a solution for the isomorphic $SRP$ where the loop-prevention mechanism for BGP is removed (i.e., we don't block paths with loops). Since we have relative-transfer-equivalence, by removing the loop condition, we get full transfer-equivalence. We can then simply appeal to Theorem~\ref{thm:correctness-local} to derive \eafwd and preference-equivalence of $\lab$ and $\ol{\lab}$ for the isomorphic networks.

    Finally, if we add back the abstract edges that we removed, we need to show that we still have the same solution $\ol{\lab}$ with loop-prevention. We do this by showing that such edges would be rejected as loops.
    Given that we have \cpequivalence, and in the concrete solution $\lab$ this edge would result in a loop of the form $u,w_1,\ldots,w_k,u$, we know that the abstract path would also have a loop $f_r(u),f_r(w_1),\ldots,f_r(w_k),f_r(u).$ 
  \end{proof}
\end{thm}

Next we show the other direction. Note that in both cases, the proof is constructive and thus responsible for identifying the particular appropriate refinement $(f_r, h_r)$ and $(f_s, h_s)$.

\begin{thm} \label{thm:refinement-equivalence2}
  Suppose we have well-formed $SRP$, $\wh{SRP}$, and $\ol{SRP}$ for BGP with an \effective abstraction $(f, h)$. For any solution $\ol{\lab}$ to $\ol{SRP}$, then there exists a refinement $(f_r, h_r) \sqsubseteq_{(f_s,h_s)} (f, h)$ where $\lab$ is a solution to $SRP$, and $\lab$ and $\ol{\lab}$ are label- and fwd-equivalent.

  \begin{proof}
    Setup $f_r$ such that, for each node $\ol{v}$ that forwards for an attribute that is not the best when ignoring loop-prevention, we have a single node in $v \in V$ map to such a $\ol{v}$. For every other node $\ol{v}$, that forwards to the best available option, we map every other $v$ to each of these $\ol{v}$. That is, we assign a single concrete node for each unique behavior that is not the best route and all other nodes map to the abstract node that has the best route.

    As before, we remove each edge in $\ol{SRP}$ that corresponds to an edge rejected due to loops in $\ol{\lab}$, and all corresponding concrete edges related under $f_r$. As before, in the concrete network, we will still have a \aaabs since each concrete node that forwards to a non-best path does so because the better paths are rejected due to loops.

    This network will have the same solution $\ol{\lab}$ but has full transfer-equivalence, and we can again appeal to Theorem~\ref{thm:correctness-local} for preference- and \cpequivalence for BGP without loop-prevention.

    If we add back the abstract edges that we removed, we need to show that the same solution $\lab$ is still a solution with loop-prevention. Suppose that the abstract node $\ol{u}$ prevented a loop $\ol{u},\ol{w}_1,\ldots,\ol{w}_k,\ol{u}$.
    Then each node $u$ where $f_r(u) = \ol{u}$ that chose a non-best path in the concrete network also did so due to loop-prevention.
  \end{proof}

\end{thm}
  
\vspace{1em}
\begin{corollary} \label{thm:bgp-equiv}
  Suppose we have well-formed $SRP$, $\wh{SRP}$, and $\ol{SRP}$ for BGP with an \effective abstraction $(f,h)$. There is a solution $\lab$, where each node $u_1 \mapsto \wh{u_1}$ forwards along path $s = \lab(u_1) \ldots \lab(u_k)$ to some node $u_k \mapsto \wh{u_k}$ \ifft there is a solution $\ol{\lab}$ where each node $\ol{u_1} \mapsto \wh{u_1}$ forwards along path $\ol{s} = \lab(\ol{u_1}) \ldots \lab(\ol{u_k})$ to some $\ol{u_k} \mapsto \wh{u_k}$ such that $h(s) = h_s(\ol{s})$.

  \begin{proof} We show each direction separately.

    \vspace{.8em}
    \noindent \textbf{Case ($\Rightarrow$)}
    Suppose $\lab$ is a solution for $SRP$. From Theorem~\ref{thm:refinement-equivalence}, we know there exists a refinement $(f_r, h_r) \sqsubseteq (f,h)$ of for $\ol{SRP}$ with solution $\ol{\lab}$, and also that $\lab$ and $\ol{\lab}$ are fwd-equivalent. 
    Given any two nodes $u$ and $v$ where $u$ can reach $v$, there exists a path $p = u,w_1,\ldots,w_k,v$ where $(u,w_1) \in \fwd(u)$ and $(w_i,w_{i+1}) \in \fwd(w_i)$ and $(w_k,v) \in \fwd(w_k)$. Because $\lab$ and $\ol{\lab}$ are fwd-equivalent, we know that $(f_r(u),f_r(w_1)) \in \ol{\fwd}(f_r(u))$ and so on. 
    Therefore, there is an abstract path in $\ol{\lab}$ where $f_r(u)$ can reach $f_r(v)$ of the form $f_r(u), f_r(w_1), \ldots, f_r(w_k), f_r(v)$. Since $f_r$ is \emph{onto} from Theorem~\ref{thm:refinement-equivalence}, we know that this is the case for every $f_r(u) \in f_s^{-1}(f(u))$. 
    Observe that the labels of the concrete path are $s = \lab(u), \lab(w_1), \ldots, \lab(w_k), \lab(v)$. Similarly, the abstract path has labels $\ol{\lab}(f_r(u)), \ldots, \ol{\lab}(f_r(v))$. 
    Due to label-equivalence, this is the same as $h_r(\lab(u)), \ldots, h_r(\lab(v))$, which is just $h_r(s)$.
    
    Recall that we must show that $h_s(\ol{s}) = h(s)$
    Since we know $h_r(s) = \ol{s}$, we have $h_s(h_r(s)) = h(s)$. Finally, because $h = h_s \circ h_r$, these are equivalent.

    \vspace{.8em}
    \noindent \textbf{Case ($\Leftarrow$)}
    Suppose $\ol{L}$ is a solution for $\ol{SRP}$, then from Theorem~\ref{thm:refinement-equivalence}, we know that  there exists an onto refinement $(f_r,h_r) \sqsubseteq (f,h)$ where $\lab$ and $\ol{\lab}$ are \eafwd. Consider an arbitrary path $\ol{u},\ol{w_1},\ldots,\ol{w_k},\ol{v}$. Then we know $(\ol{u},\ol{w_1}) \in \ol{\fwd}(\ol{u})$ and so on. 
    From the fact that $\lab$ and $\ol{\lab}$ are \eafwd, every node $u$ that maps to $\ol{u}$ forwards to the same neighbor. That is, we know that each $u$ where $f_r(u) = \ol{u}$, has the same $w_1$ where $f_r(w_1) = \ol{w_1}$ and $(u,w_1) \in \fwd(u)$ and so on. 
    Therefore, each node $u \in f_r^{-1}(\ol{u})$ has the same path starting after $w_1$: $u,w_1,\ldots,w_k,v$. The abstract path has labels $\ol{\lab}(f_r(u)),\ldots,\ol{\lab}(f_r(v))$. Due to label-equivalence, this is the same as $h_r(\lab(u)),\ldots,h_r(\lab(v))$, which is $h_r(s)$. 
    
    Once again, we have $h_s(h_r(s)) = h(s)$, which follows from the fact that $h = h_s \circ h_r$.
  \end{proof}
\end{corollary}

\end{document}